\documentclass[reprint, superscriptaddress, amssymb, aps, pra, a4]{revtex4-2}
\usepackage{graphicx}
\usepackage{subcaption}
\usepackage{dcolumn}
\usepackage{hyperref}
\usepackage{braket}
\usepackage{bm}
\usepackage[utf8]{inputenc}
\usepackage{natbib}
\usepackage{amsmath}
\usepackage{tabularx}
\usepackage{xcolor}
\usepackage[english]{babel}
\usepackage{soul}
\usepackage{siunitx}
\usepackage{comment}
\usepackage[normalem]{ulem}
\newcommand{\p}{\mathrm{p}}
\newcommand{\s}{\mathrm{s}}
\renewcommand{\i}{\mathrm{i}}
\renewcommand{\d}{\mathrm{d}}

\newcommand{\A}{\,\hat{a}^{\dagger}}

\newcommand{\sugg}[1]{\textcolor{blue}{#1}}

\begin{document}

\title{Advances in Position-Momentum Entanglement: \\ A Versatile Tool for Quantum Technologies}

\author{Satyajeet Patil}
\affiliation{Institute for Applied Physics, Technical University of Darmstadt, Schloßgartenstraße 7, 64289 Darmstadt, Germany}
\author{Sebastian Töpfer}
\affiliation{Institute for Applied Physics, Technical University of Darmstadt, Schloßgartenstraße 7, 64289 Darmstadt, Germany}
\author{Rajshree Swarnkar}
\affiliation{Institute for Applied Physics, Technical University of Darmstadt, Schloßgartenstraße 7, 64289 Darmstadt, Germany}
\author{Sergio Tovar-Perez}
\affiliation{Institute for Applied Physics, Technical University of Darmstadt, Schloßgartenstraße 7, 64289 Darmstadt, Germany}
\author{Jonas Moos}
\affiliation{Institute for Applied Physics, Technical University of Darmstadt, Schloßgartenstraße 7, 64289 Darmstadt, Germany}
\author{Jorge Fuenzalida}
\affiliation{Institute for Applied Physics, Technical University of Darmstadt, Schloßgartenstraße 7, 64289 Darmstadt, Germany}
\affiliation{Present Address: ICFO-Institut  de  Ciencies  Fotoniques,  The  Barcelona  Institute  of Science  and  Technology,  08860  Castelldefels  (Barcelona),  Spain}
\author{Markus Gräfe}
\affiliation{Institute for Applied Physics, Technical University of Darmstadt, Schloßgartenstraße 7, 64289 Darmstadt, Germany}

\date{\today}

\begin{abstract}
Position-momentum entanglement is a versatile high-dimensional resource in quantum optics. From fundamental tests of reality to applications in quantum technologies, spatial entanglement has experienced significant growth in recent years. In this review, we explore these advances, beginning with the generation of spatial entanglement, followed by various types of measurements for certifying entanglement, and concluding with different quantum-based applications. We conclude the review with a discussion and outlook of the field.

\end{abstract}


\maketitle

\section{Introduction}\label{S1}
\subsection{Background}\label{S1A}

In 1935, Einstein, Podolsky and Rosen (EPR) raised the question of the completeness of the quantum mechanical description of reality~\cite{EPR}. Theoretically, EPR showed that, if there are two non-separable, nonlocal correlated systems, then measuring non-commutative physical quantities (for example, position and momentum) of the first system reveals the position and momentum of the second system precisely, without even disturbing it. Indeed, this was counter-intuitive to the Heisenberg uncertainty principle (HUP). So they stated that HUP is the consequence of the incompleteness of the quantum mechanical description and not the fundamental principle by itself. In such non-separably correlated systems, reality of the second system depends upon the measurements carried out on the first system without disturbing it in any way even if the systems are light-like separated from each other, no reasonable definition of reality could be expected to permit this. This is famously known as the EPR paradox. Einstein called this new nonlocal quantum mechanical phenomenon as \textit{spooky action at a distance} and Schrödinger coined a term called \textit{entanglement} for it \cite{Schrödinger_1935}. Later in 1957, Bohm and Aharonov presented such a correlated system of a molecule with total spin zero consisting of two atoms, each with spin one-half \cite{Bohm1957}. The aim of their work was to present a clearer and experimentally accessible version of the EPR paradox that can be tested in the laboratory. EPR suggested that there should be some real parameters known as \textit{local hidden variables} which can determine the outcomes of the entangled states. The parameter is called hidden variable because it was not directly accessible to experiment. According to EPR, this parameter was missing from the quantum mechanical description. John Bell in 1964 delved into comparative study of hidden variable theory and quantum mechanics. He formulated a theorem which concludes that no local hidden variable theory could reproduce all the predictions of quantum mechanics~\cite{Bell1964}. He derived an inequality, a mathematical condition that puts an upper bound on the correlations governed by local hidden variables, i.e. $\mid$S$\mid \leq 2$, where $S$ is known as Bell's parameter. In contrast, quantum mechanics allows stronger correlations, with a maximum value known as the Tsirelson bound, which is $|S| \leq 2\sqrt{2}
$ ~\cite{Tsirelson1980}. 
To enable the experimental testing of Bell’s theorem under realistic conditions, John Clauser, Michael Horne, Abner Shimony, and Richard Holt reformulated Bell’s original inequality to account for specific experimental constraints, resulting in the widely used CHSH inequality~\cite{CHSH}: $2<$ $\mid$S$\mid$ $\leq$ $2\sqrt{2}$. Under specific conditions, entangled quantum states violate this bound. These studies were crucial because they allowed us to distinguish between quantum-mechanical predictions and local hidden-variable predictions. These tests confirm that quantum mechanics is a complete theory with no need of hidden variables. The measurement of Bell's parameter has paved the way for exploring different quantum mechanical tests, including numerous practical applications.\\
Bell’s parameter serves as a tool to demonstrate quantum nonlocality, which in turn implies the presence of entanglement beyond what can be explained by classical theories. Nowadays, Bell theorem is a useful resource
in different quantum technological applications. Some quantum key distribution (QKD) protocols, e.g. Ekert91 protocol~\cite{PhysRevLett.67.661}, use Bell's inequality to ensure secure communication. Quantum teleportation~\cite{QuantTelBoumeester} utilizes entangled states which are often validated by Bell's parameter. Quantum random number generation using entangled states are quantified using Bell's parameter~\cite{Jacak2020,PhysRevLett.120.010503}. Entanglement based quantum metrology is dependent on the quality of entanglement that is characterized using Bell's parameter~\cite{Matthews2016}.\\
However, entanglement could also be generated in continuous-variables, such as transverse position-momentum \cite{SpaEntHowell, imaging_certifying_entanglement, PhysRevA.85.013827, Achatz2020Certifying,Zhang:19, PATIL2023128583}, energy-time \cite{PhysRevLett.62.2205, xavier2025energytime, shalm2013three,ETAli2,ETKwiat2, ETkwiat1,ETAli1}, positionlike-momentumlike quadratures \cite{ou1992quantum,quadrature_microwave_entanglement,zhang_experimental_2015} and so on. To evaluate entanglement, there are several criteria such as EPR criterion~\cite{SpaEntHowell, PhysRevLett.90.043601, edgar2012imaging,PhysRevLett.113.160401,OAMEntLeach}, Renyi entropy~\cite{PhysRevA.95.042334, PhysRevA.83.032307}, and partial transpose~\cite{PhysRevLett.87.167904,PhysRevLett.84.2726}. Among these criteria, EPR criterion is extensively used to differentiate between classical and quantum systems. Violation of EPR criterion, validates quantum nonlocality/entanglement.\\
So, mathematically, the EPR criterion for the entangled state in position-momentum follows,\cite{Zhang:19, edgar2012imaging,PhysRevLett.113.160401}

\begin{equation}
\Delta^{2}\left(x_2 \mid x_1\right) \Delta^{2}\left(p_{2 x} \mid p_{1 x}\right) < \frac{\hbar^{2}}{4} ,
\label{E1}
\end{equation}
\\
where $\Delta^{2}\left(x_2 \mid x_1\right)$ denotes the minimum inferred variance in the transverse position of photon 2, given knowledge of photon 1’s position, and $\Delta^{2}\left(p_{2 x} \mid p_{1 x}\right)$ denotes the minimum inferred variance in the transverse momentum of photon 2, given knowledge of photon 1’s momentum. \\
 The EPR criterion of entanglement has a wide range of applications, especially where continuous-variable entanglement is exploited. As discussed before, the EPR criterion provides a mathematical framework to test nonlocality. Such tests probe into entanglement and provide insights of into the fundamental nature of reality. The EPR criterion has been shown to be useful in continuous-variable quantum key distribution (CV-QKD), where phase- or amplitude-modulated coherent or squeezed states of quadrature variables are used to encrypt and transmit information \cite{scarfe_quantum_2025}. Quantum teleportation of continuous-variables using EPR entangled states is often validated by the EPR criterion~\cite{PhysRevLett.80.869}. Wide-field quantum imaging extensively relies on transverse spatial correlation, either in momentum or position, whereas the correlation width settles the resolution limit in some techniques, more details in Sec.~\ref{resolution-section}. The smaller the value of the EPR criterion, the stronger the entanglement and better the resolution ~\cite{defienne_quantum_2019, Defienne2022PixelSuperResolution, defienne_full-field_2021, Reichert2017, Ndagano2022QuantumMicroscopy,moreau2018resolution,fuenzalida2022resolution,viswanathan2021resolution,vega2022fundamental,gilaberte2023experimental}. Entangled states satisfying EPR criterion are used for quantum metrology for enhanced sensing and measurements beyond the classical limit~\cite{PhysRevLett.96.010401,Lyons2016Precision,DiLorenzo2013Correlations, Bullock2014Focusing}. Beyond the idealized formulations, generalized EPR states have been explored to better represent experimentally relevant biphoton wavefunctions, and various metrics have been proposed to quantify their entanglement and nonclassical features \cite{adhikari2025characterizing}.

\subsection{Spontaneous parametric down-conversion}\label{S1B}

\begin{figure}
    \centering
    \includegraphics[width=0.4\textwidth]{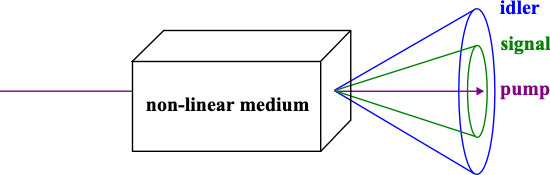}
    \caption{In a SPDC process, a high-energy pump photon (purple) interacts with a second-order nonlinear crystal to generate two lower-energy photons, called signal (green) and idler (blue), emitted at characteristic angles. The emission angles of the SPDC photons depend on the phase-matching conditions.}
    \label{F1}
\end{figure}

Bell's theorem motivated to look for different sources of entangled states of particles/photons. In the early days, some of the methods to realize entangled states were spin-zero molecules (as given by Bohm \cite{Bohm1957}), radiative atomic cascade of calcium to create polarization entangled biphoton states \cite{Alain, Alain1982, Kocher}, and non-linear optical processes. In recent years, a second-order non-linear process known as spontaneous parametric down conversion (SPDC) has become a widely used method to create correlated photon pairs. In the SPDC process, a higher energy photon interacts with the non-linear medium and emits two lower energy photons, historically known as signal and idler (see Figure~\ref{F1}). The SPDC process was first theoretically predicted by Louisell \textit{et al.} in 1961 \cite{FirstSPDCTheory}, who
modeled it as two oscillators coupled via a time-dependent
reactance. Later, Giallorenzi \textit{et al.} \cite{SPDCscatteringGiallorenzi} explained SPDC as the scattering process in which a pump photon is transformed into a photon pair because of the non-linear polarization. Its first experimental observation was studied by Harris \textit{et al.} in 1967~\cite{FirstSPDCexpt}, later followed by many others. Burnham \textit{et al.} in 1970 experimentally confirmed simultaneous emission of signal and idler photons via coincidence detection~\cite{FirstCCdetectionBurnham}. The coincidence-detection study paved the way for use of SPDC photon pairs for fundamental study \cite{hanburybrown1956test} as well as the application based explorations.\\ 
\begin{figure}
    \centering
    \includegraphics[width=0.4\textwidth]{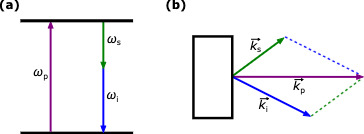}
    \caption{Conservation relations of the (a) energy and (b) momentum. } 
    \label{F2}
\end{figure}
Boyd \textit{et al.}~\cite{boyd_nonlinear_2008} explained SPDC as a result of the second-order nonlinear polarization in non-centrosymmetric crystals. Assuming the absence of resonant atomic excitation and pump field strengths much smaller than the atomic field strength, this process will follow the laws of energy and momentum conservation in lossless media, Figure~\ref{F2} shows this schematically. Because of energy conservation, the total energy of the generated photons equals the energy of the pump photon. Therefore
\begin{equation}
    \mathbf{\omega}_\p = \mathbf{\omega}_\s + \mathbf{\omega}_\i,
    \label{E2}
\end{equation}
with $\omega_\mathrm{p,s,i}$ as the angular frequency of the pump, signal and idler photons, respectively. Equally, the momentum conservation laws are fulfilled with, 
\begin{equation}
    \mathbf{k}_\p = \mathbf{k}_\s + \mathbf{k}_\i\,,
        \label{E3}
\end{equation} 
where $\mathbf{k}_\mathrm{p}$, $\mathbf{k}_\mathrm{s}$, and $\mathbf{k}_\mathrm{i}$ are the wavevectors of the pump, signal, and idler, respectively. As can be seen in Figure~\ref{F2}(b), to fulfill the momentum conservation, the transversal components of the momentum vectors are anti-correlated between the signal and idler photons. Both photons originate from the same positions in the crystal, leading to a direct correlation in the transversal position. Furthermore, the generated photons show correlation in various degrees of freedom such as polarization \cite{PolEntRubin, PolEntKwiat1, PolEntKwiat3, PolEntKwiat2}, orbital angular momentum (OAM) \cite{OAMEntLeach, OAMEntRarity,OAMEntMair}, energy-time \cite{ETAli2, ETKwiat2, ETkwiat1, ETAli1}, and transverse position and momentum \cite{SpaEntHowell, law2004analysis, SpaEntWolborn}

\begin{figure}
    \centering
    \includegraphics[width=0.4\textwidth]{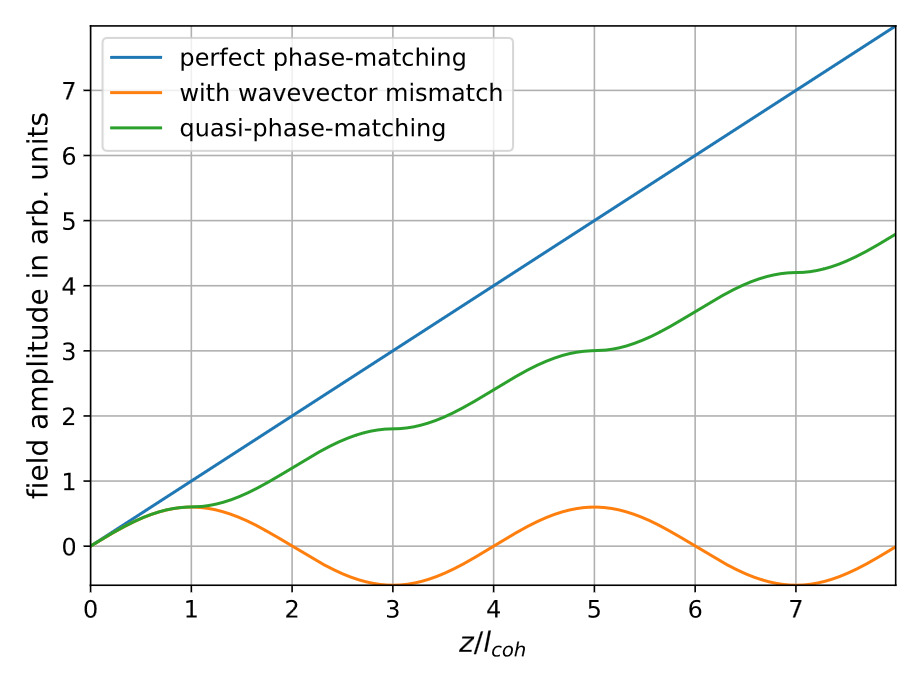}
    \caption{Amplitude and phase-matching conditions. The generated SPDC light amplitude depends on the propagation length $z$ and the coherent build-up $l_\textbf{coh}$ of the conversion process for (blue) no wavevector mismatch, (orange) an existing wavevector mismatch and (green) quasi-phase matching. Reproduced from Ref.~\cite{boyd_nonlinear_2008} with permission from Elsevier.}
    \label{F3}
\end{figure}

In order to achieve efficient SPDC, it is additionally necessary to set the longitudinal momentum mismatch

\begin{equation}
    \Delta \mathbf{k}_z = \mathbf{k}_{\mathrm{p}z} - \mathbf{k}_{\mathrm{s}z} - \mathbf{k}_{\mathrm{i}z},
        \label{E4}
\end{equation}
\\
with $\mathbf{k}_z$ as the longitudinal component of the wavevector set to zero. This is called phase matching. As visualized in Figure~\ref{F3} there will be a linear dependence of the propagation length with the field amplitude of the generated SPDC signal, if the condition is fulfilled (perfect phase matching). In case of $\Delta \mathbf{k}_z \neq 0$, the field amplitude will vary periodically depending on the propagation length as a consequence of the generated light waves that are destructively interfering. Phase matching is dependent on a multitude of parameters, e.g the involved wavelengths, the crystal's internal structure, the temperature and more. Therefore, the possible wavelength combinations for perfect phase matching are limited in number. In the year 1962 Armstrong \textit{et al.} \cite{armstrong_interactions_1962} proposed to periodically modulate the sign of the second-order nonlinear coefficient of the material to shift the phase of the generated light waves. This method is called quasi-phase matching and is a versatile tool for increasing the possible spectral combinations of SPDC. As a consequence, the phase-matching condition for the quasi-phase-matching case changes to

\begin{equation}
    \Delta \mathbf{k}_z \equiv 0 = \mathbf{k}_{\mathrm{p}z} - \mathbf{k}_{\mathrm{s}z} - \mathbf{k}_{\mathrm{i}z} - \frac{2 \pi}{\Lambda} , 
    \label{E5}
\end{equation}
\\
with $\Lambda$ as the period of the alternating second-order nonlinear coefficient.

\subsection{Discrete variable entanglement}\label{S1C}
In the case of polarization, when phase matching condition is such that generated photons have the same polarization as the pump photon then it is known as type-0 phase matching. When generated photons have orthogonal polarization to that of pump photon then it is known as type-I phase matching and when the generated photons have orthogonal polarization with respect to each other then it is known as type-II phase matching. The polarization correlation between the signal and idler have been extensively studied for fundamental tests as well as the quantum communication and information applications. The study of polarization correlation really exploded when Aspect et al. showed strong violation of Bell's inequality for the first time in 1981 \cite{Alain}. In their experiment, they studied the linear polarization correlation of the photons emitted in the radiative atomic cascade of calcium. This remarkable study ruled out the whole class of realistic local theories and became a foundational stone for many non-classical tests and applications. In recognition of these foundational contributions, Alain Aspect, Anton Zeilinger, and John Clauser were awarded the 2022 Nobel Prize.

However, the atomic cascade source is difficult to handle and has low brightness. To overcome this issue, Kwiat \textit{et al.} used an SPDC source consisting of BBO (beta barium borate) with type-II phase matching \cite{PolEntKwiat3}. With improved visibility of 97\% they demonstrated the violation of Bell's inequality. They also mentioned a method to convert one Bell state into another (Equation\ref{E6}) using unitary transformations.
\begin{equation}
\begin{aligned} 
    \left|\psi^{ \pm}\right\rangle &= \frac{1}{\sqrt{2}} \left(\left|H_\s, V_\i\right\rangle \pm\left|V_\s, H_\i\right\rangle\right), \\
    \left|\phi^{ \pm}\right\rangle &= \frac{1}{\sqrt{2}} \left(\left|H_\s, H_\i\right\rangle \pm\left|V_\s, V_\i\right\rangle\right).
\end{aligned}
\label{E6}
\end{equation}

Generation of Bell states played a crucial role in demonstrating many quantum mechanical phenomena like quantum teleportation\cite{QuantTelBoumeester}, super-dense coding\cite{QuantDenseAnton}, entanglement-based quantum communication\cite{BBM92,Mishra_2022}, quantum metrology\cite{PhysRevLett.96.010401, Giovannetti2011Advances}, no-cloning theorem\cite{wootters1982nocloning}, etc. 

\par
In addition to linear polarization, light also possesses circular polarization. Circularly polarized photon\sugg{-}states are the eigenstates of the spin angular momentum (SAM) operator with eigen values $\pm\hbar$. This was first demonstrated experimentally by Beth in 1936 \cite{SAMBeth}. Photons with linear polarization are associated with zero SAM. Unlike SAM, OAM of light is consequence of the helical wavefront and azimuthally varying phase \cite{LOAM,Raskatla:22,Bekshaev:20}. OAM carrying beam has a phase singularity at the center of the beam and flow of energy spirals around the propagation axis hence it is famously known as an optical vortex (OV). Optical vortices carry \( \ell\hbar \) OAM per photon \cite{Kopf2025}, where \( \ell \) is the order of orbital angular momentum. Spin angular momentum spans two-dimensional space, whereas OAM spans infinite-dimensional space ranging from \( -\infty \) to \( +\infty \).

Along with the energy and momentum, conservation of orbital angular momentum is also maintained in SPDC processes. Making use of this fact, Mair \textit{et al.} \cite{OAMEntMair} for the 
 first time showed entanglement generation of photons in the OAM degree of freedom. This work encouraged various other groups to explore the properties
of OAM in the quantum domain \cite{OAMMolina-Terriza2007} for information transfer and processing \cite{OAMfickler}. Just as polarization of photons represents a qubit, OAM states of light have also been used to represent higher-dimensional bits known as qudits. Orbital angular momentum states of light provide infinitely many orthogonal basis states that can be used to encode quantum information and may improve security \cite{Anwar:21,Karimipour2002,Bechmann-Pasquinucci2000}. Controlled generation of OAM entangled state is studied for secure quantum communication and information processing \cite{OAMManagementof,OAMExperimentalVaziri,OAMPreparationTorres, OAMAnwar_2020, OAMJabir_2019}. \\

\subsection{Continuous-variable entanglement}\label{S1D}
Photon pairs produced in a SPDC process can also be correlated/entangled in continuous-variables such as quadrature phase-amplitude\cite{ou1992quantum,quadrature_microwave_entanglement,zhang_experimental_2015}, position-momentum entanglement \cite{SpaEntHowell, imaging_certifying_entanglement, PhysRevA.85.013827, Achatz2020Certifying,Zhang:19, PATIL2023128583}. In recent years, the field of continuous-variable entanglement study \cite{SpaEntHowell} has been explored extensively, yielding significant progress in both theoretical frameworks and experimental implementations. In \cite{PhysRevLett.60.2731}, Reid and Drummond derived the EPR criterion to show how it could be utilized with position-like and momentum-like quadrature observables of squeezed states to show inseparability/entanglement. Shortly after that, \cite{PhysRevLett.68.3663, PhysRevLett.84.2722, PhysRevLett.84.2726} derived sufficient and necessary conditions for continuous-variable entanglement. Transverse correlations of the photons produced in the SPDC process have been studied both theoretically as well as experimentally  \cite{law2004analysis,strekalov_observation_1995, pittman_optical_1995, PhysRevA.57.3123, PhysRevA.63.063803}. 

\par
Transverse position and momentum are the spatial variables of the signal and idler, which satisfy EPR criterion. Howell \textit{et al}. \cite{SpaEntHowell} were the first to report strong position-momentum entanglement in the photon pairs generated in the SPDC process. Using coincidence detection in the near and far field, they measured position-momentum variance product of 0.01$\hbar^{2}$, which is direct indication of violation of the bounds for the EPR and separability criteria. The spatial entanglement is continuous-variable entanglement and has gained attention in recent years, and found applications in the field of quantum imaging \cite{defienne_quantum_2019, Defienne2022PixelSuperResolution, defienne_full-field_2021, Reichert2017, Ndagano2022QuantumMicroscopy}, quantum metrology \cite{PhysRevLett.96.010401, Giovannetti2011Advances} and quantum communication \cite{Achatz2020Certifying, scarfe_quantum_2025, PhysRevLett.96.090501, PhysRevA.77.062323}. As discussed in the Sec.~\ref{S2}, conditional uncertainty/correlation length of the position and momentum of the two photons depends on several pump and non-linear crystal parameters. The momentum correlation length depends on the beam waist and the spatial coherence of the pump, whereas, the position correlation length depends on the crystal length and the wavelength of the pump beam. Several authors have shown the effect of spatial coherence on the degree of spatial entanglement \cite{Zhang:19, Huggo, Giese2018} whereas a few have investigated the effect of different spatial profiles of pump beam on entanglement \cite{PATIL2023128583, Boucher:21, verniere_hiding_2024}. Such studies have been found useful for tuning the spatial entanglement, for comparing the position-momentum correlation length of the SPDC photons created using coherent laser and the incoherent LED, shaping the angular spectrum of the SPDC photons etc. Several other studies delve into propagation dynamics of spatial entanglement \cite{PhysRevA.95.063836, AbhiProagation}.

\section{Theoretical Background}\label{S2}
The biphoton quantum state in the wave number representation at the crystal face (z=0) reads \cite{PhysRevA.57.3123, chan2007trasnverse, PhysRevA.79.033801}
\begin{equation}
    |\Psi(z)\rangle=\int \d \mathbf{k}_{\s} \d \mathbf{k}_{\i} \Phi(\mathbf{k}_{\s}, \mathbf{k}_{\i}, z) \A_s(\mathbf{k}_{\s}) \A_\i(\mathbf{k}_{\i})\ket{0,0}\,,
    \label{E7}
\end{equation}
\\
where $\Phi(\mathbf{k}_{\s}, \mathbf{k}_{\i}, z)$ is known as the mode function. $\mathbf{k}_{\s}$ and $\mathbf{k}_{\i}$ are the wavevectors of the signal and idler photons and $\A_s(\mathbf{k}_{\s}), \A_\i(\mathbf{k}_{\i})$ are the corresponding creation operators. Under the condition of collinear phase matching condition (pump, signal and idler propagating in the same direction), $\Phi(\mathbf{k}_{\s}, \mathbf{k}_{\i}, z)$ is given by, 

\begin{equation}
\begin{aligned}
\Phi({\mathbf{k}}_{\s}, {\mathbf{k}}_{\i}, z)= & N E_\p \operatorname{sinc}\left(\frac{\Delta{\mathbf{k}}_z L}{2}\right) \exp \left(\i \frac{\Delta{\mathbf{k}}_z L}{2}\right) \\
& \times \exp \left\{\i\left[{\mathbf{k}}_{\s z}+{\mathbf{k}}_{\i z}\right] z\right\},
\end{aligned}
\label{E8}
\end{equation}

where \textit{N} is the normalization factor, $E_{p}$ is the transverse profile of the pump beam at the crystal plane ($
E_p \equiv \exp\!\left(-\frac{w_p^{\,2}}{4}\,\bigl|\mathbf{k}_{s}+\mathbf{k}_{i}\bigr|^2\right)$) , and $L$ is crystal length. $\Delta{\mathbf{k}}_{z} = \mathbf{k}_{\p z} - \mathbf{k}_{\s z} - \mathbf{k}_{\i z}$ is the phase mismatch in the $z$-direction. The sinc function describes the efficiency of down-conversion as a function of phase mismatch. When phase mismatch is zero, i.e. $\Delta \mathbf{k}_{z} =0$, sinc function is maximized ($\mathrm{sinc}(0)=1$). The value of the sinc function decreases as the phase mismatch increases, which in turn reduces the efficiency of the down conversion. $\exp \left\{\i\left[\mathbf{k}_{\s z}+\mathbf{k}_{\i z}\right] z\right\}$ represents the phase evolution of the signal and idler photons propagating in the $z$-direction. $\exp \left(\i \frac{\Delta \mathbf{k}_{z} L}{2}\right)$ represents the phase shift that arises from the phase mismatch $\Delta \mathbf{k}_{z}$ within the non-linear crystal.

\par
For the collinear phase matching condition, the sinc function in Equation~\eqref{E8} can be approximated to a Gaussian function by using an adjustment parameter $\alpha = 0.455$. Moreover, if we consider that the angular spectrum of the pump beam $E_p$ is given by a Gaussian function, then, the momentum space wave function of the bi-photons at the crystal plane takes the form as~\cite{edgar2012imaging, PhysRevA.79.033801}\\

\begin{equation}
\begin{aligned}
\Psi\left( \mathbf{k}_\s,  \mathbf{k}_\i\right)=\frac{\sigma_{+} \sigma_{-}}{\pi} \exp \left[-\frac{\left|\mathbf{k}_\s+ \mathbf{k}_\i\right|^2}{4 \sigma_{+}^2
}\right]\\
\times \exp \left[-\frac{\sigma_{-}^2}{4}\left| \mathbf{k}_\s- \mathbf{k}_\i \right|^2\right].
\label{E9}
\end{aligned}
\end{equation}
\\
Here, $\sigma_{\pm}$ are the standard deviations of the two Gaussian functions and are related to the correlation widths.
$\sigma_{-}$ represents the position correlation width of the bi-photons, and it is defined as
\begin{equation}
\begin{aligned}
\sigma_{-} = \sqrt{\frac{\alpha \, L  \, \lambda_p}{2 \pi}}, 
\end{aligned}
\label{E10}
\end{equation}
\\
where $L$ is the crystal length, and $\lambda_p$ is the pump beam's wavelength. $\sigma_{+}$ is the momentum correlation width and reads 
\begin{equation}
\begin{aligned}
\sigma_{+} = 1/(2 \, w_p),
\label{E11}
\end{aligned}
\end{equation}
which is the inverse of the pump waist $w_p$.
\par
By applying a Fourier transform to Equation~\eqref{E9}, it is obtained the biphoton state in position space
\begin{equation}
\begin{aligned}
\Psi\left(\mathbf{x}_\s, \mathbf{x}_\i\right)=\frac{1}{\pi \sigma_{-} \sigma_{+}} \exp \left[-\frac{\sigma_{+}^2}{4}\left|\mathbf{x}_\s+\mathbf{x}_\i\right|^2\right]\\
\times \exp \left[-\frac{\left|\mathbf{x}_\s-\mathbf{x}_\i\right|^2}{4 \sigma_{-}^2}\right].
\label{E12}
\end{aligned}
\end{equation}
\par
If the pump waist is much greater than the position correlation width, i.e., $\sigma_{-} \ll \sigma_{+}^{-1} $, then Eqs.~\eqref{E9} and \eqref{E12} indicate that the signal and idler photons will exhibit position correlation and momentum anti-correlation. The effect of the various pump beam parameters on these correlations is discussed in detail in Secs.~\ref {S5} and~\ref{S6}. \\

\section{Techniques to measure spatial entanglement}\label{S3}
In general, position-momentum entanglement can be measured with various techniques using different resources and assumptions on the quantum system. These techniques can be categorized as entanglement estimation, entanglement certification, and entanglement quantification~\cite{friis2019entanglement}. In entanglement estimation, a parameter roughly obtains the spatial resources of a system. One example is the number of spatial modes extracted from measuring the pump waist and the properties of the crystals~\cite{moreau2018resolution, brambilla2004simultaneous,kviatkovsky2020microscopy,hong2024polarization}. Next is entanglement certification, which can often be achieved by measuring intensity profiles in the far-field plane (assuming a certain purity of the quantum state). For example, one can use the Schmidt number ~\cite{law2004analysis,lorenzo2009direct,ortega2022spatial} to certify spatial entanglement by comparing intensity profiles in the near vs. far field, or by measuring the coherence of the field ~\cite{just2013transverse,Bhattacharjee_2022}. See also~\cite{PhysRevA.84.063847, Kulkarni2017, PhysRevA.97.063846} for similar approaches.
Finally, entanglement quantification is the most rigorous approach, measuring the full two-photon correlation both in near field and far field. Because it is so comprehensive, a number of advanced measurement techniques have been developed to realize it experimentally. Entanglement quantification methods include: (i) direct position and momentum detection of signal/idler photons using movable slits [example refs]; and (ii) intensity-correlation measurements with imaging sensors (EMCCD cameras or single-photon detector arrays). Here, we focus on entanglement quantification techniques detailing their experimental realizations.  

\subsection{Two-photon coincidences in the near and far field}\label{S3C}

\par
\begin{figure}[htbp]
    \centering
    \includegraphics[width=0.4\textwidth]{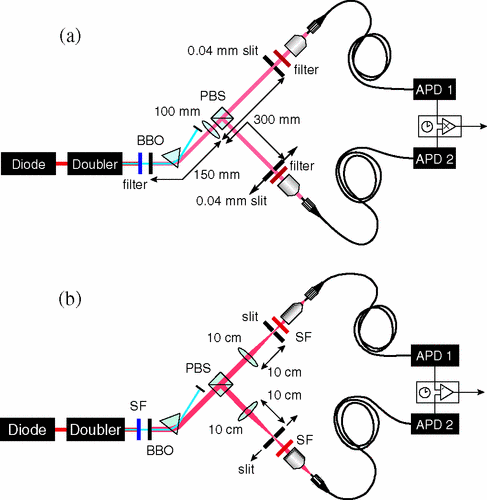}
    \caption{Experimental setups for measuring (a) position correlation width and (b) momentum correlation width. Reproduced with permission. \cite{SpaEntHowell} Copyright © 2004, American Physical Society.}
    \label{F4}
\end{figure}

From this section onward, we discuss methods that are based on correlation measurements at near- and far-field.
Ref.~\cite{SpaEntHowell}, reports an experimental realization of the EPR paradox using position-momentum entangled photons. In the experimental setup, a 2~mm long type-II BBO crystal was pumped with a 390~nm pump beam to generate SPDC photon pairs. A prism spectrally cleans the down-converted photons from the pump beam. The orthogonally polarized signal and idler are separated using a polarizing beam splitter (PBS) and collected at two different detectors. Additionally, in each detector, a slit of width 40 $\mu$m is placed to clean the modes spatially. To measure the correlation widths of position and momentum, lenses are placed to form the near-field and far-field configurations in Figs.~\ref{F4}(a) and ~\ref{F4}(b), respectively. For measuring position correlations, the crystal plane is imaged onto the slit; whereas for measuring momentum correlations, the crystal plane is Fourier-transformed onto the slit. Behind each slit, there is a $10$~nm band-pass filter followed by a coupling system which is connected to a coincidence counter.
 In both cases, one of the slits (slit-1) is fixed at the peak of single-photon counts, while the other slit (slit-2) is free to move along the $x$-axis. Coincidence count rates are recorded as a function of displacement of slit-2, giving the distributions for position and momentum. By normalizing coincidence distribution, the conditional probability $P(x_2|x_1)$ for position and $P(p_2|p_1)$ for momentum are
 obtained. These probability distributions are then used to calculate the uncertainty/standard deviation in position and momentum. The experimentally obtained values are $\Delta(x_{2}|x_{1})=0.027$~mm and $\Delta(p_{2}|p_{1})=3.7 \, \hbar$ mm$^{-1}$. The product of the variances for position and momentum violates the separability bound by two orders of magnitude

\begin{equation}
    \Delta^{2} (x_{2}|x_{1})\Delta^{2} (p_{2}|p_{1}) = 0.01 \hbar^2.
    \label{E15}
\end{equation}

Theoretically, for the given crystal length, pump beam's wavelength and waist, the variance should be $0.0036\, \hbar^{2}$, which differs from the experimental value. The report suggests this discrepancy is perhaps due to different experimental imperfections. Additionally, it should be noted that, because of the broader slit width (40 $\mu m$) conditional probability distribution is broader than the one associated with the SPDC process by itself. 

\par
\begin{figure}[htbp]
    \centering
    \includegraphics[width=0.45\textwidth]{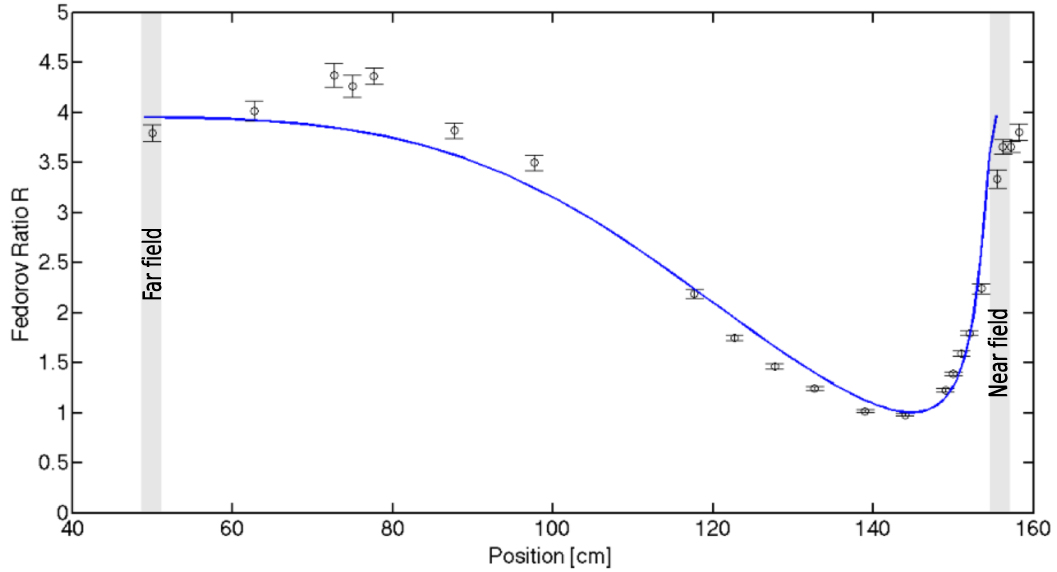}
    
\caption{Fedorov ratio measured between the near-field and far-field. Reproduced from \cite{just2013transverse}. Copyright © 2013 IOP Publishing Ltd.}

    \label{F6}
\end{figure}

Noteworthy to mention an additional technique that quantifies the spatial entanglement by measuring intensities and correlations in both planes: the Fedorov ratio~\cite{just2013transverse, brida2009characterization,federov2004packet}. In it, the intensity profiles are divided by the conditional probability of the biphotons in far-field and near-field planes. When the Fedorov ratio equals 1, no entanglement is present in the system. However, this measurement technique fails when measured between the near field and the far field, as the value of 1 does not represent a system without entanglement. The reason for this is that the spatial entanglement is transferred from the amplitude to phase while it moves from the far field to the near field, see Figure~\ref{F6}. More details in Refs. \cite{chan2007trasnverse,brida2009characterization}.

\begin{figure}
    \centering
    \includegraphics[width=0.3\textwidth]{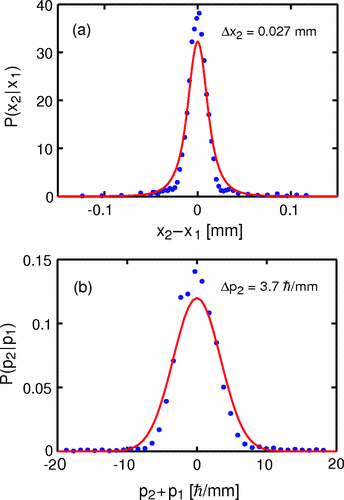}
    \caption{Distribution of coincidence counts against the displacement of the
slit-2 for (a) position and (b) momentum. Reproduced with permission. \cite{SpaEntHowell} Copyright © 2004, American Physical Society.}
    \label{F5}
\end{figure}

\subsection{Correlation imaging using a camera} \label{S3D}
\par
Although coincidence counting with slits provides a robust way to certify spatial entanglement, it is limited by low photon collection efficiency. As a result, modern experiments increasingly rely on high-sensitivity imaging devices to capture spatial entanglement more efficiently. With the advancement
in detection technology, a weak signal can be detected and registered with quantum
efficiency $>$90\% using an electron-multiplying CCD (EMCCD) camera. The general scheme to extract correlation using an EMCCD camera uses the sub-Poissonian statistics of photon pairs and
is described in details in \cite{PhysRevLett.120.203604}. Several other authors have also used EMCCD cameras and 2D detection arrays for measuring spatial entanglement \cite{imaging_certifying_entanglement, PATIL2023128583, edgar2012imaging, Defienne2022PixelSuperResolution, PhysRevLett.120.203604, Eckmann:20, Kundu:24, Courme:23, 10059123}. In simple form, the schematic of this technique is depicted in Figs.~\ref{F8} and {\ref{F9}. A nonlinear crystal is pumped by a continuous wave laser of wavelength 405 nm. A band-pass filter removes the pump beam from the setup and transmits the generated SPDC photons at 810 nm. For near-field/position correlations, a lens of focal length $f$=50 mm images the crystal plane onto the EMCCD sensor with a magnification $M$=1. For far-field/momentum correlations, a lens of focal length of $f$=100 mm Fourier images the crystal plane onto the EMCCD sensor. In this way, a plane wave carrying a wavevector $\mathbf{k}$ is focused to a point given by a pixel of the camera. As depicted in Figure~\ref{F7}, one of the photons from the photon pair hits the EMCCD camera sensor at pixel \textit{i} (while the other lands at pixel \textit{j}), which is then converted into photoelectrons with a quantum efficiency of $\eta$. These photoelectrons are then amplified in a process that allows the weakest signal to become readable. Then, amplified electrons are converted into grey-scale values corresponding to pixel \textit{i}(\textit{j}). Although the measurement process can be optimized \cite{PhysRevA.98.013841}, the certification of spatial entanglement with an EMCCD camera still requires millions of images to work.

\par
To measure the spatial correlations, the joint probability distribution (JPD) is again calculated in near and far-field configurations, but in terms of the camera pixels. The JPD $\Gamma (r_i,r_j)$ represents the probability of detecting one photon from a photon pair at the pixel \textit{i} and its partner at the pixel \textit{j}, thus, \cite{PATIL2023128583, defienne_quantum_2019,Defienne2022PixelSuperResolution, Huggo, PhysRevLett.120.203604}

\begin{equation}
\Gamma(r_i, r_j)
= \frac{1}{N} \sum_{l=1}^{N} I_l(r_i) I_l(r_j)
- \frac{1}{N^{2}} \sum_{\substack{l,\, l' \\ l \neq l'}}^{N} I_l(r_i) I_{l'}(r_j).
\label{E16}
\end{equation}

\textit{N} is the total number of frames, $I_l(r_i)$ is the intensity at the pixel $r_i$ from an frame number $l$, similarly for pixel $r_j$. The first term on the right side is the average tensorial product between the intensities of a frame $l$ with itself. 
Similarly, the second term is the tensorial product of the frame $l$, but with the next frame, $l^{'}$ (or $l+1$). Since SPDC photons are created simultaneously and reach the image sensor in an exact moment in time, the second term will only contribute with accidental coincidences. By subtracting the second and first terms, only genuine coincidences from photon pairs are obtained, see Figure~\ref{F10}.

The standard deviations of the joint conditional probabilities are given by \cite{PATIL2023128583} 

\begin{equation}
    \Delta(x_{\s}|x_{\i}) = \frac{\sigma_{x(\text{emccd})}}{M}, \Delta(p_{x{\s}}|p_{x{\i}}) = \frac{k \sigma_{p(\text{emccd})}}{ \, f}
    \label{E17}
\end{equation}
\\
where $\sigma_{x(\text{emccd})}$ is the position correlation width, $M$ is the magnification, $\sigma_{p(\text{emccd})}$ is the momentum correlation width, $k$ is the wavenumber of the SPDC photons, and $f$ is the focal length of the lens used for the Fourier image.
\begin{figure}
    \centering    \includegraphics[width=0.4\textwidth]{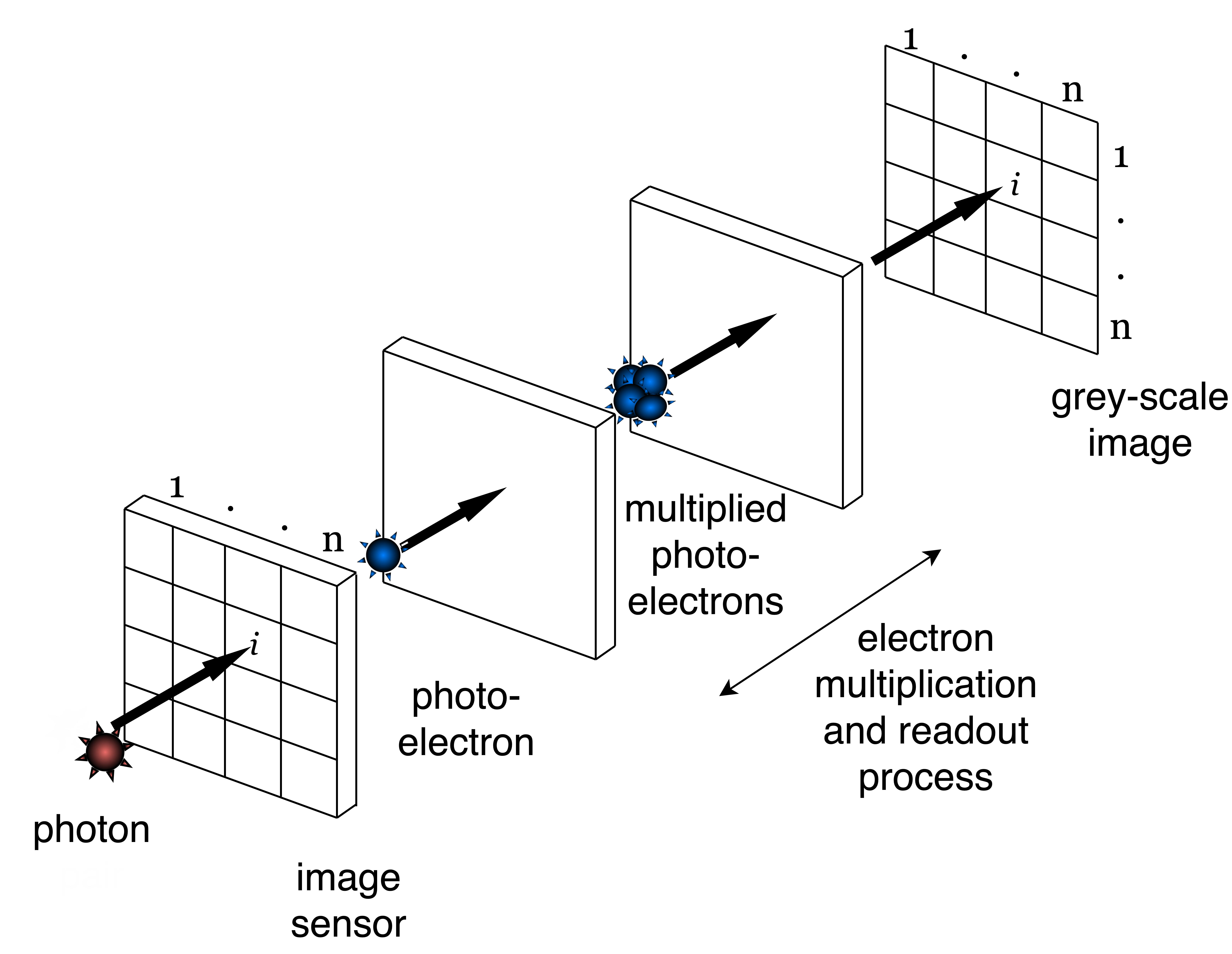}
    \caption{Illustration of the process by which a single photon is converted into an amplified signal in an EMCCD camera, demonstrating the key stages of photon absorption, electron multiplication, and signal readout.}
    \label{F7}
\end{figure}

\begin{figure}[htbp]
    \centering
\begin{subfigure}[b]{0.55\textwidth}
         \centering
\includegraphics[width=1\linewidth]{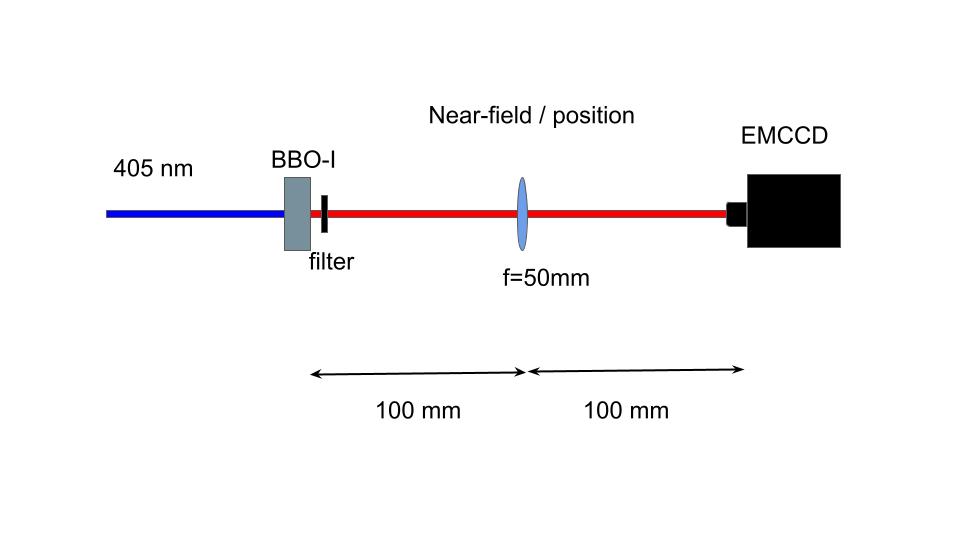}
\caption{Experimental setup for measuring the near-field (position-space) distribution of down-converted photons. Photon pairs generated via SPDC in a BBO crystal by a 405 nm pump beam are spectrally filtered and imaged onto an EMCCD camera using a lens of focal length f=50 mm, enabling direct measurement of transverse position correlations at the crystal output plane.}
\label{F8}
\end{subfigure}   
     \vfill
\begin{subfigure}[b]{0.55\textwidth}
         \centering
\includegraphics[width=1\linewidth]{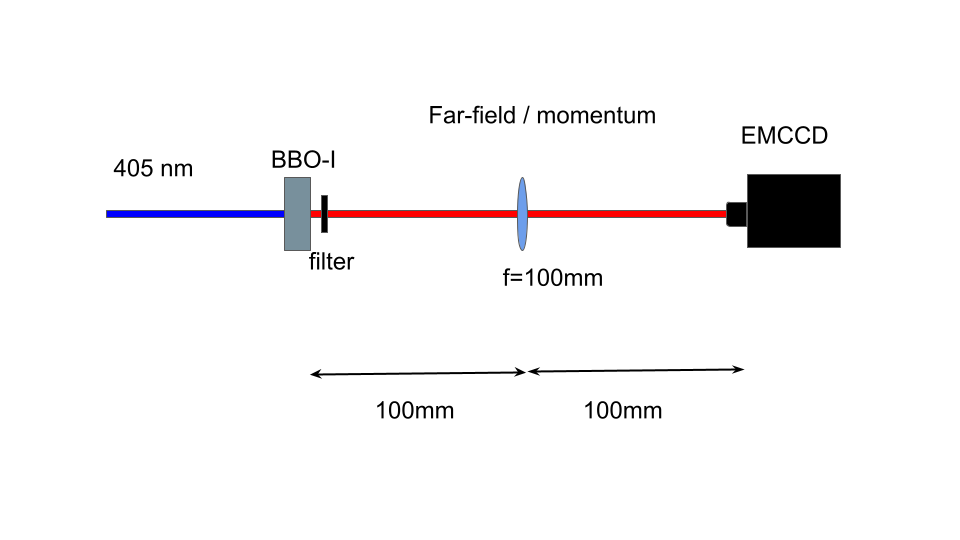}
\caption{Experimental setup for measuring the far-field (momentum-space) distribution of down-converted photons. A 405 nm pump beam generates photon pairs via type-I SPDC in a BBO crystal. After spectral filtering, a lens of focal length f=100 mm is placed such that the EMCCD camera is located in the Fourier plane, enabling detection of transverse momentum correlations.}
\label{F9}
\end{subfigure}
\caption{Experimental setups to certify spatial entanglement using an EMCCD camera.} 
\end{figure}

\par

To visualize the photon pairs' correlations in different configurations, one sums the frame's rows or columns as shown in Figure~\ref{F11}.
These summations transform the $(n\times n)$ matrix into a $(n \times 1)$ row vector (or $(1\times n)$ if columns are summed) which has the information of intensity correlation in the $x$-axis ($y$-axis). Multiplying the transpose of the row vector with the row vector within the same frame $l$, the pixel-to-pixel intensity correlation for the x-axis is obtained, see Figure~\ref{F10}. Likewise, to get the second JPD in the y-axis, this process is repeated for the column vector. 

\begin{figure}[h!]
    \centering
\includegraphics[width=0.55\textwidth]{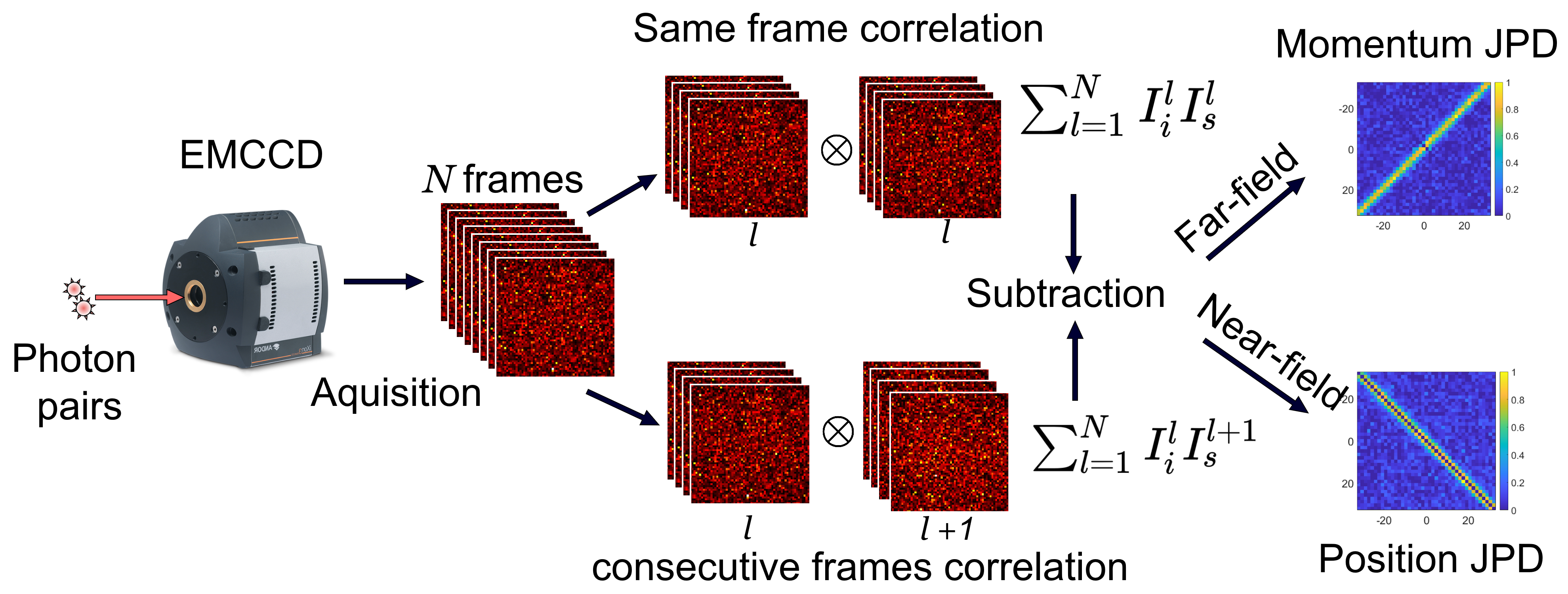}
    \caption{Illustration of the process for obtaining the joint probability distribution of momentum and position. Photon pairs are recorded using an EMCCD camera over millions of images. Correlations within the same frame include both genuine and accidental coincidences, while correlations between consecutive frames contain only accidental coincidences. By subtracting the consecutive frame correlation from the same frame correlation, genuine coincidences are isolated. The frames are then summed along the rows or columns to obtain the correlations along the x- or y-axes. }
    \label{F10}
\end{figure}

\begin{figure}
    \centering
    \includegraphics[width=0.55\textwidth]{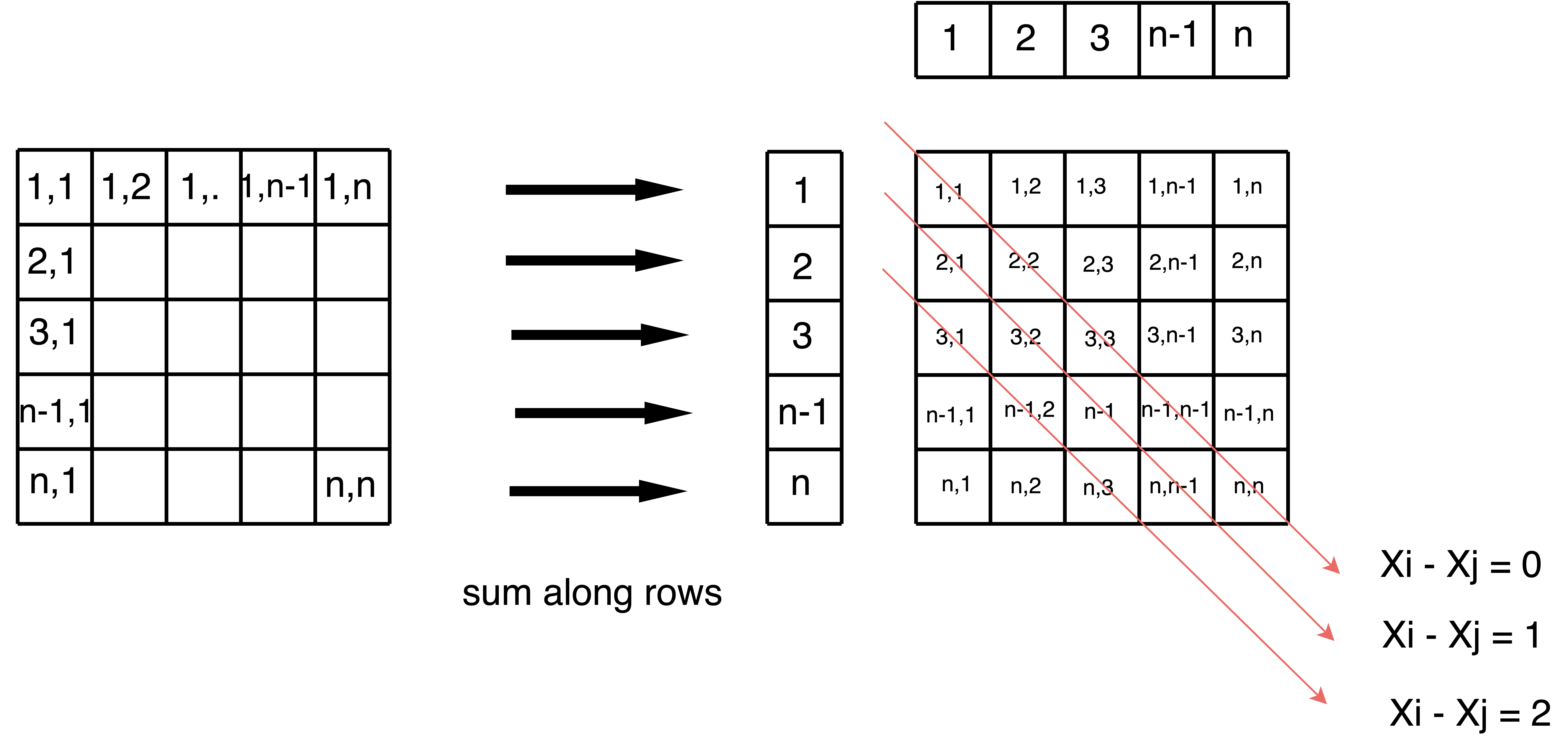}
    \caption{Illustration of the process where an image is summed along rows to obtain a column vector. This column vector is then multiplied by its transpose, resulting in an $n \times n$ matrix. In the resulting matrix, diagonal elements represent intensity correlation for the same pixel, the first sub-diagonal elements correspond to intensity correlation between consecutive pixels, and so on.}
    \label{F11}
\end{figure}
\par

The JPD is achieved through statistical averaging, roughly over $10^6$ to $10^7$ intensity images. Given the fact that CCD-based detectors provide frame rates of the order of 100 frame per second, the total acquisition time to achieve JPD can vary from several hours to over a day. Such long acquisition times hinder the adaptation of quantum imaging schemes for real-world applications. However, new camera technologies can overcome these drawbacks. 

In recent years, single-photon avalanche diode (SPAD) cameras have been developed~\cite{10059123, 1206776} and are now commercially available. SPAD arrays have demonstrated their capabilities in LiDAR (Light detection and ranging)~\cite{6872788,8502386,8662355}, in fluorescence lifetime imaging~\cite{Bruschini2020, Li:10,8494330}, in imaging through strongly scattering media~\cite{Lyons2019}, in non-line of sight imaging~\cite{Gariepy2016}, in time-resolved correlation measurements~\cite{Lubin:19,Unternahrer:16}, and also, to certify spatial entanglement~\cite{imaging_certifying_entanglement}. The detection principle in SPAD cameras allows the detection of joint events of photon pairs between different pixels. This fundamental difference, compared to other cameras, enables entanglement certification in just seconds, with a significantly improved signal-to-noise ratio. With this, SPAD cameras have shown potential for real-world applications in quantum technologies. But, despite all these advantages, SPAD cameras still need time to become a leading technology, for example, to achieve higher resolution and lower prices.

\section{Effect of spatial coherence on entanglement} \label{S5}

This section examines how the spatial coherence of the pump beam influences the degree of position    –momentum entanglement of the photon pairs. To understand how the coherence of the pump beam influences spatial correlations, the theoretical study by Jha and Boyd \cite{Jha2010Spatial} provides valuable insight, while the experimental investigation by Zhang \textit{et al.} \cite{Zhang:19} offers complementary verification. In the experimental setup shown in Figure~\ref{F13}, in the blue-shaded part of the setup, they employ an LED to serve as a spatially incoherent light source, whereas in the red-shaded part, they employ a coherent laser followed by a spatial light modulator (SLM) that manipulates the transverse phase profile of the light. A nonlinear crystal is used to generate SPDC, and finally, a PBS is used to guide the correlated photon pairs into individual arms.

The authors used the method described in Sec.~\ref{S3C} to measure the near- and far-field JPDs of the SPDC photons emitted from the LED and the laser.
Their experimental results are shown in Figure~(\ref{F14}). The  transverse positions of signal and idler photons are highly correlated in both cases, for the laser in Figure~\ref{F14}(a) and for the LED in Figure~\ref{F14}(c). This indicates that the position correlations are not influenced by the pump beam's coherence, but rather by other factors, see equation~\eqref{E10} for details. 
On the contrary, the momentum anti-correlations are influenced by the pump beam's coherence. The results show that a sharper distribution for the momentum is obtained when a laser beam is employed, see Figure~\ref{F14}(b), in comparison to when an LED is used, Figure~\ref{F14}(d)}. 

The broadening of the momentum correlations can be explained as follows. Considering a Gaussian-Schell model for the partial spatially coherent beam~\cite{RevModPhys.37.231}, and a Gaussian approximation for the down-converted field ~\cite{Huggo, Fedorov_2009, Schneeloch_2016}, the momentum correlation width reads

\begin{equation}
    \sigma_k = \sqrt{\frac{1}{l_c^2} + \frac{1}{4w_p^2}}\text{\,}.
    \label{E29}
\end{equation}

$l_c$ is the coherence length of the pump beam and $w_p$ is the pump waist. When a perfectly coherent pump beam is considered ($l_c\to\infty$), equation \eqref{E29} reduces to equation \eqref{E11}.

\begin{figure}[hbtp]
    \centering
    \includegraphics[width=0.5\textwidth]{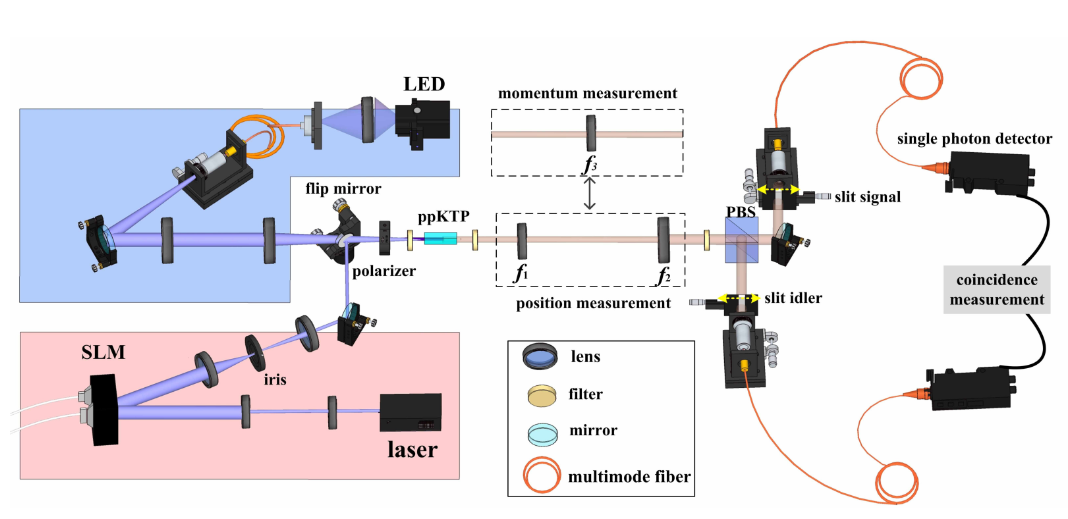}
    \caption{Experimental setup for the characterization of the joint spatial distributions of the photon pairs. Position correlations are measured in the image plane ($f_1$ and $f_2$), while momentum correlations are determined in the Fourier plane ($f3$). Reproduced from Ref.~\cite{Zhang:19} under the terms of the Creative Commons Attribution (CC BY) license.
}
    \label{F13}
\end{figure}

\begin{figure}[ht]
    \centering
    \includegraphics[width=0.35\textwidth]{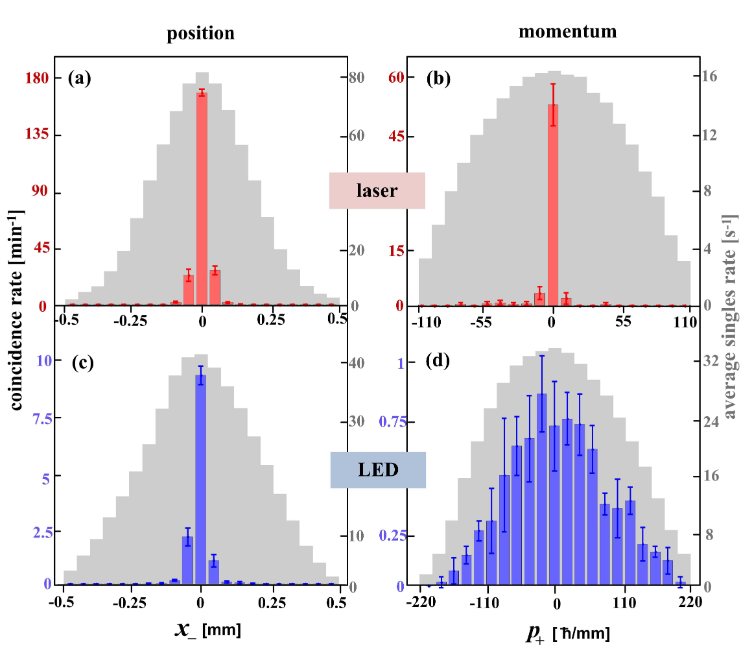}
    \caption{Position and momentum correlations of SPDC using laser and LED pumping. (a), and (c) shows the coincidence rates that are recorded in the near field to obtain position distributions. (b) and (d) are the coincidence rates measured in the far field for momentum distributions. Reproduced from Ref.~\cite{Zhang:19} under the terms of the Creative Commons Attribution (CC BY) license.}
    \label{F14}
\end{figure}

\begin{figure}[ht]
    \centering
    \includegraphics[width=0.35\textwidth]{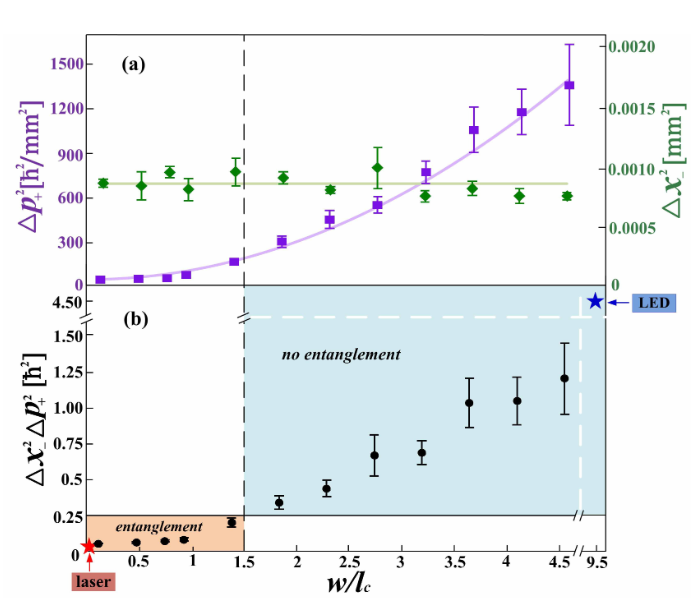}
    \caption{Momentum anti-correlation, position correlation, and entanglement relationship for pump beams with varying coherence lengths. Reproduced from Ref.~\cite{Zhang:19} under the terms of the Creative Commons Attribution (CC BY) license.}
    \label{F16}
\end{figure}

Evidently, the broadening in momentum correlations also has consequences for the spatial entanglement. Figure~\ref{F16} shows different values related to the spatial entanglement using different experimental parameters. The red star, corresponding to the SPDC generated by the laser, exhibits high entanglement, whereas the blue star, corresponding to the SPDC generated by the LED, displays classical correlations. This way, Figure~\ref{F16} shows the physical system transitioning from entangled to classically correlated photon pairs caused by the broadening of the momentum correlations. Similarly, parallel work by Defienne \textit{et al.}~\cite{Huggo} reported consistent findings.

While reduced pump coherence is generally associated with weaker momentum correlations and diminished spatial entanglement in SPDC photon pairs, a recent work has demonstrated that this relationship is not universal. Under specific conditions, partially coherent or structured pump beams can, in fact, enhance spatial entanglement. Hutter \textit{et al.} \cite{Hutter2020Boosting} demonstrated that illuminating a nonlinear crystal with a Twisted Gaussian–Schell model (TGSM) beam---a partially coherent beam with a controllable twist phase---can strengthen position–momentum correlations, contrary to the usual expectation. The twist phase introduces a coupling between transverse position and momentum variables in the pump, leading to biphoton states that exhibit stronger nonclassical correlations even as the pump coherence decreases.
This finding overturns the conventional assumption that coherence and entanglement scale monotonically. In the infinite-dimensional spatial Hilbert space, mixed states can remain highly entangled even at low purity. Quantitative analysis revealed that decreasing the coherence length or increasing the twist phase reduces the biphoton purity but simultaneously increases its entanglement, as verified through symplectic eigenvalue analysis. These results highlight that incoherent or structured illumination can be exploited as a resource for tailoring and enhancing entanglement in SPDC systems, broadening the experimental conditions under which strong spatial correlations can be achieved. Such techniques open promising possibilities for quantum imaging.

\section{Engineering Joint Probability Distribution}\label{S6}

In this section, we discuss methods to actively shape or engineer the joint position–momentum probability distribution of the SPDC photon pairs. To tailor the shape of the JPD of the down-converted photons, one can directly act on the photon pair \cite{PATIL2023128583,Boucher:21,Lib:20,cameron2024tutorial} or, as the pump's angular spectrum is transferred to the SPDC photons \cite{PhysRevA.57.3123,lib2}, such pump beam can be used. Methods that act directly on the SPDC photon pairs include wave-front shaping~\cite{PhysRevLett.121.233601,Hugospdcshaping}, quantum interferometry~\cite{Ferreri_2020,yingwen}, metasurfaces~\cite{kaiwang,tomer}, and rotating diffusers~\cite{Lib2022}.
In the following, we present how the influence of the pump beam's spatial profile and phase matching condition, influences the JPD of the down-converted photons.

\begin{figure}[hbtp]
    \centering
    \includegraphics[width=0.4\textwidth]{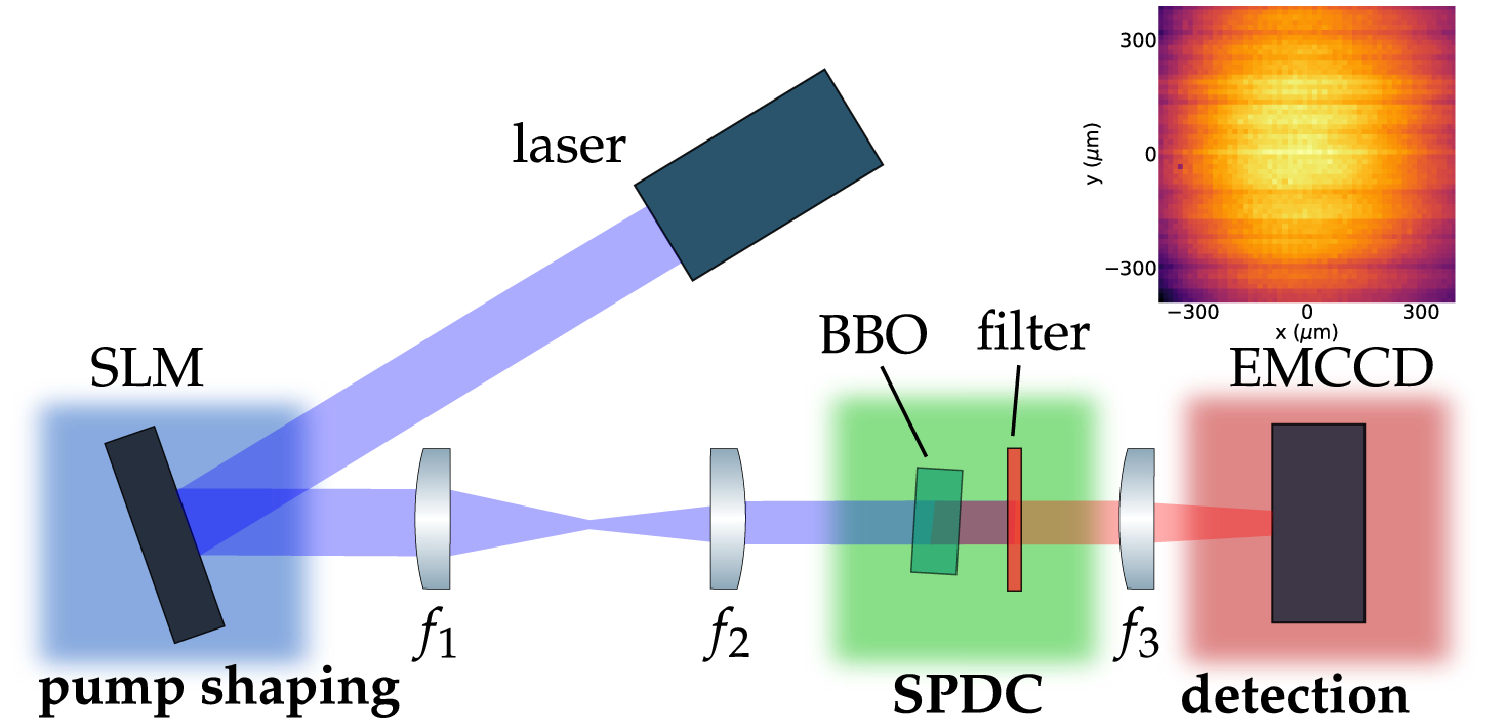}
    \caption{Schematic of the experimental setup. A SLM is used to shape the pump beam, which is then imaged onto the crystal plane using the lenses $f_{1}$ and $f_{2}$. The lens $f_3$ Fourier images the crystal plane onto the EMCCD plane. Reproduced (Adapted) with permission. \cite{Boucher:21} Copyright © 2021, Optica.
}
    \label{F25}
\end{figure}

Reference~\cite{Boucher:21} presents a general study of the spatial shaping of the pump beam with a spatial light modulator (SLM). The experimental setup is depicted in Figure~\ref{F25}. A continuous-wave laser illuminates the SLM, which tailors its beam profile, and subsequently, impinges on a type-I nonlinear crystal using a 4-f imaging system. To get the angular spectrum of the generated SPDC photon-pairs, a lens with focal length $f=d/2$ is inserted between the crystal and the EMCCD camera, where $d$ is the distance between the camera and the nonlinear crystal.\\ 

Although many parameters of the SPDC process can be tuned to modify the spatial correlations, the authors decided to focus on the case of a Bessel-Gaussian beam. This beam can be described as~\cite{Boucher:21}

\begin{equation}
B G(r, \phi, 0 ; l)=A J_l\left(k_r r\right) \exp (i l \phi) \exp \left(-\frac{r^2}{w_g^2}\right),
\label{E37}
\end{equation}
\par
where A is the constant, $J_l$ is the $l^{th}$ Bessel function of the first kind, $w_g$ is the beam waist of the Gaussian component, $k_r$ is the radial spatial frequency, and $(r,\phi)$ is a pair of polar coordinates. The angular spectrum---i.e., the Fourier transform of the Bessel-Gaussian beam in Equation~\eqref{E37}---reads

\begin{equation}
\begin{aligned}
& \mathcal{V}_p^{B G}\left(k, \phi_k, 0 ; l\right)= \\
& \quad i^{l-1} \frac{w_g}{w_0} \exp \left(i l \phi_k\right) \exp \left(-\frac{k^2+k_r^2}{w_0^2}\right) I_l\left(\frac{2 k_r k}{w_0^2}\right),
\end{aligned}
\label{E38}
\end{equation}
\par
where $w_0=2/w_{g}$, and $I_l$ is the $l^{th}$ order modified Bessel function of the first kind. From this, the two-photon state generated using zeroth-order Bessel-Gaussian pump beam is given as

\begin{equation}
\begin{aligned}
&\Phi\left(\mathbf{k}_\s, \mathbf{k}_\i\right)= \frac{\pi \delta}{k_r} \frac{-i w_g^2}{\i} \exp \left(-4 \frac{\left|\mathbf{k}_\s+\mathbf{k}_\i\right|^2+k_r^2}{w_g^2}\right) \\
& \qquad \times I_0\left(\frac{8 k_r\left|\mathbf{k}_\s+\mathbf{k}_\i\right|}{w_g^2}\right) \exp \left(-\frac{\delta^2\left|\mathbf{k}_\s-\mathbf{k}_\i\right|^2}{2}\right).
\end{aligned}
\label{E39}
\end{equation}

\begin{figure}[hbtp]
    \centering
    \includegraphics[width=0.45\textwidth]{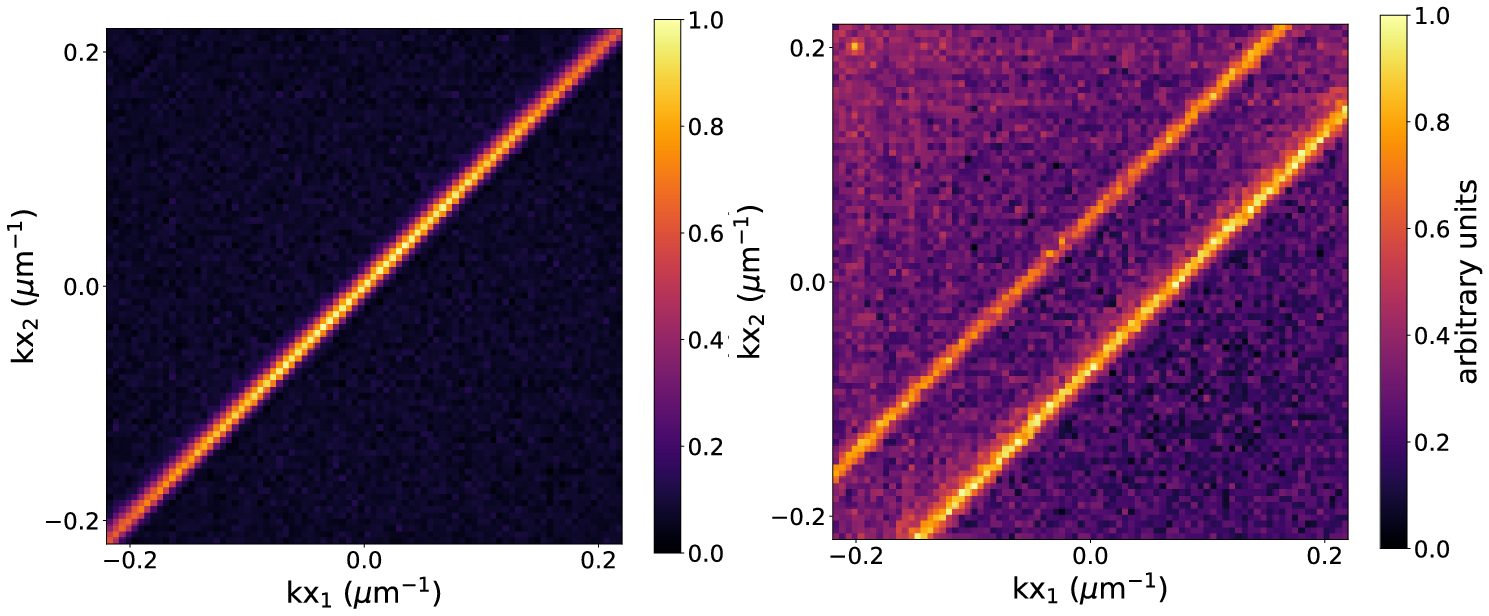}
    \caption{ Photon pair's JPD in the far-field plane. (Left) JPD obtained with a Gaussian pump beam, and (right) the JPD obtained with a Bessel-Gaussian pump beam. Reproduced (Adapted) with permission. \cite{Boucher:21} Copyright © 2021, Optica.}
    \label{F27}
\end{figure}

The JPD ($\Gamma (k_\s,k_\i)$) is calculated using the method discussed in Sec.~\ref{S3D}. Figure~\ref{F27} shows the SPDC anti-correlation in momentum for Gaussian (left) and Bessel-Gaussian (right) pump beams. For a Bessel-Gaussian pump beam, the anti-correlation splits into two well-defined lines. These results clearly demonstrate that spatial correlations can be engineered by pump manipulation. In particular, tuning the angular spectrum of the pump leads to results in the far-field plane.

In another work, Ref.~\cite{PATIL2023128583}, the authors investigated the controlled engineering of spatial entanglement in SPDC by shaping the pump beam into an elliptical Gaussian profile, using cylindrical lenses. This introduced different beam widths along the $x$- and $y$-directions, which directly affected the momentum correlation of the photon pairs while leaving the position correlation unchanged. As the pump asymmetry increased, entanglement and the number of spatial modes along the $y$-direction decreased, whereas the entanglement along the $x$-direction remained nearly constant. For highly asymmetric pumping, entanglement in the $y$-direction vanished while strong entanglement persisted along the $x$-direction, producing anisotropic entangled states. Such tunable, direction-dependent entanglement may be useful for applications in quantum communication and quantum imaging where different correlation strengths are required.

Recent advances have highlighted the interplay between structured light and quantum entanglement, particularly in the position–momentum domain. Prasad \textit{et al}. \cite{Prasad2024Structured} demonstrated the experimental generation of two-photon structured position–momentum entanglement using SPDC. By manipulating the phase-matching conditions---either by tuning the phase-matching angle or by employing multiple nonlinear crystals---they achieved spatially structured biphoton correlations while maintaining entanglement in the system, verified through the entanglement-of-formation criterion. This work motivates the idea that spatial structuring can enable advanced imaging techniques beyond classical limits, while also opening new directions for metrology and information encoding.

\section{Propagation of Spatial entanglement}\label{S7}

This section explores the evolution of spatial entanglement in free-space propagation. 
Understanding the dynamics of the spatial entanglement is of paramount importance for several quantum technological applications. A clear example is the turbulence present in free-space quantum communications, horizontal and vertical links. By understanding the primary contributors to turbulence, typically represented by Zernike modes, methods to optimize the quantum channel can be incorporated. Furthermore, this can lead to the development of additional solutions to decoherence, such as distillation or quantum memories. An interesting case is the \textit{propagation-induced revival of angle-OAM entanglement} in Ref.~\cite{AbhiProagation}, where entanglement initially decays during free-space propagation, but as the photons travel further, the entanglement is restored. We revise this work next.

Let us consider a Gaussian beam with a pump waist $w_p(z=0)$=$w_0$ at the crystal plane. Then, the two-photon wavefunction in the position basis at a distance $z$ from the crystal is given by, 
\begin{align} 
&\Psi\left(\mathbf{x}_{\s}, \mathbf{x}_{\i}, z\right) ={} A \, \exp \left[ -\frac{\left|\mathbf{x}_{\s} + \mathbf{x}_{\i}\right|^2}{4 \, w_p(z)^2} \right] \notag \\
& \qquad \times \exp \left[ -\frac{\left|\mathbf{x}_{\s} - \mathbf{x}_{\i}\right|^2}{4 \, \sigma(z)^2} \right]
\, e^{\i \phi\left(\mathbf{x}_{\s}, \mathbf{x}_{\i}, z\right)},
\label{E40}
\end{align}

where A is a normalization constant, $\mathbf{x}_{\s}$ and $\mathbf{x}_{\i}$ are the transverse positions of signal and idler at the position $z$, respectively. $w_p(z)=w_0 \sqrt{1+z^2 /\left(k^2 w_0^4\right)}$, $ \sigma(z)=\sigma_0 \sqrt{1+z^2 /\left(k^2 \sigma_0^4\right)}$, where $k$ is the wavenumber of the pump beam. Moreover, $\sigma_0$ is the same as in Equation~\eqref{E10} at $z=0$, i.e., the crystal plane, and $e^{\i \phi\left(\mathbf{x}_{\s}, \mathbf{x}_{\i}, z\right)}$ is a phase factor. At a distance of $z$, the two-photon probability distribution function, $P(\mathbf{x}_{\s}, \mathbf{x}_{\i}; z)$ = $|\Psi\left(\mathbf{x}_{\s}, \mathbf{x}_{\i}, z\right)|^2$, reads

\begin{equation}
P\left(\mathbf{x}_{\s}, \mathbf{x}_{\i}, z\right)=|A|^2 \exp \left[-\frac{\left|\mathbf{x}_{\s}+\mathbf{x}_{\i}\right|^2}{2 w_p(z)^2}\right] \exp \left[-\frac{\left|\mathbf{x}_{\s}-\mathbf{x}_{\i}\right|^2}{2 \sigma(z)^2}\right].
\label{E41}
\end{equation}

As discussed early, the standard deviation of this probability distribution gives the conditional uncertainty $(\Delta(x_{\s}|x_{\i}))$ at a distance $z$. Alternatively, one can obtain the conditional uncertainty in the momentum basis $(\Delta(p_{x\s}|p_{x\i}))$ by applying a  Fourier transform to Equation~\eqref{E40}. In a similar manner, the conditional uncertainty in the angle, i.e., $\Delta(\theta_\s|\theta_\i )$, can also be estimated. This last parameter plays a key role in the entanglement certification, since by changing the measurement basis, entanglement can be restored.

\begin{figure}[hbtp]
    \centering
    \includegraphics[width=0.45\textwidth]{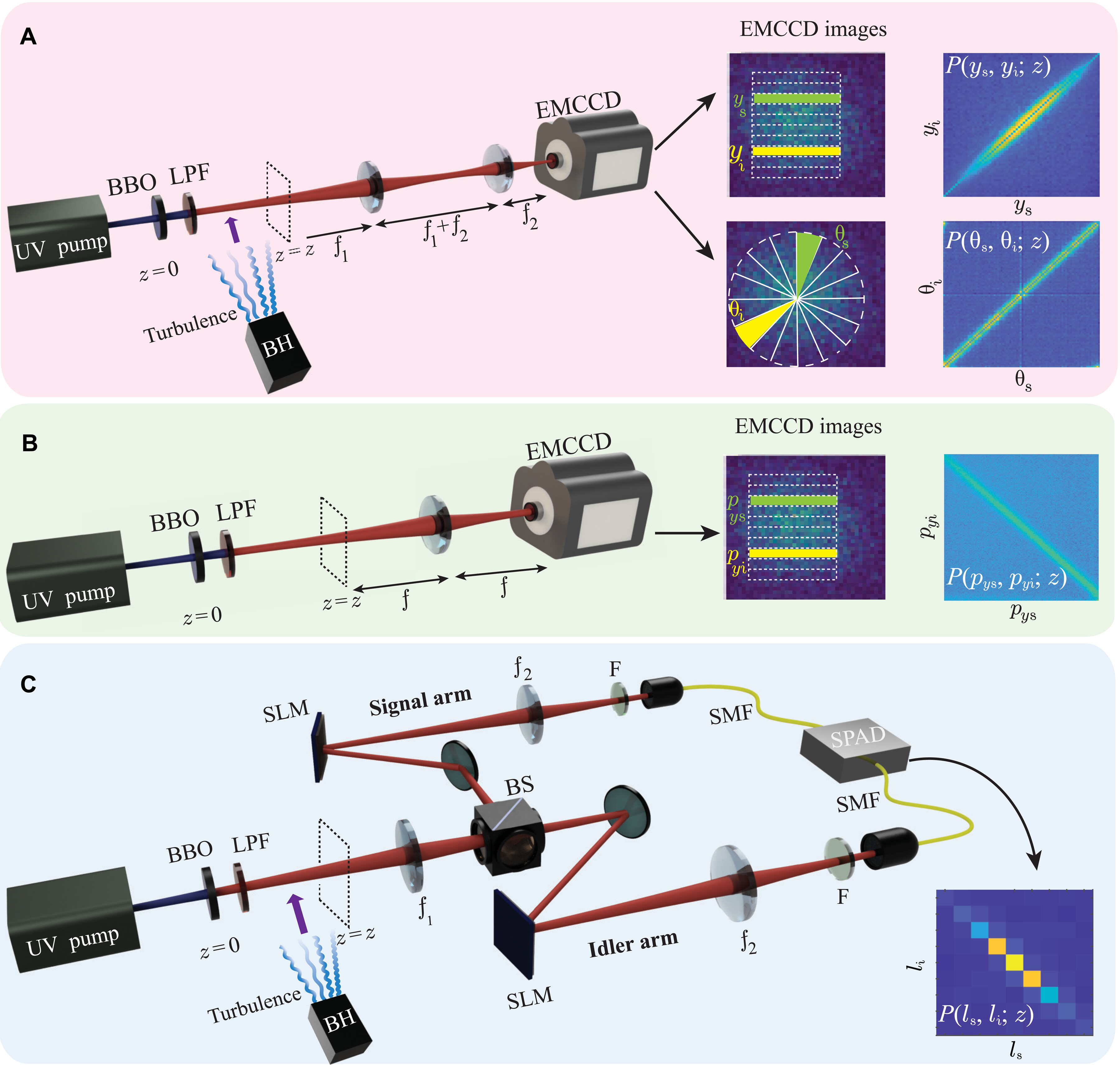}
    \caption{(A) Experimental setup to measure the conditional probability for position and angle at distance $z$ from the crystal. (B) Experimental setup to measure the momentum conditional probability and (C) to measure the OAM conditional probability. Reproduced from Ref.~\cite{AbhiProagation} under the Creative Commons Attribution license (CC BY).
}
    \label{F28}
\end{figure}

Different configurations to measure position-angle, momentum, and OAM probability distributions are shown in Figure~\ref{F28}. The main difference with previous examples presented in this review for position and momentum configurations is that the imaging systems do not act over the crystal plane, but over an arbitrary plane located at a position $z$ from the source. In addition, the OAM correlations at different $z$-positions are measured in different bases with the help of an SLM. 

\begin{figure}[hbtp]
    \centering
    \includegraphics[width=0.5\textwidth]{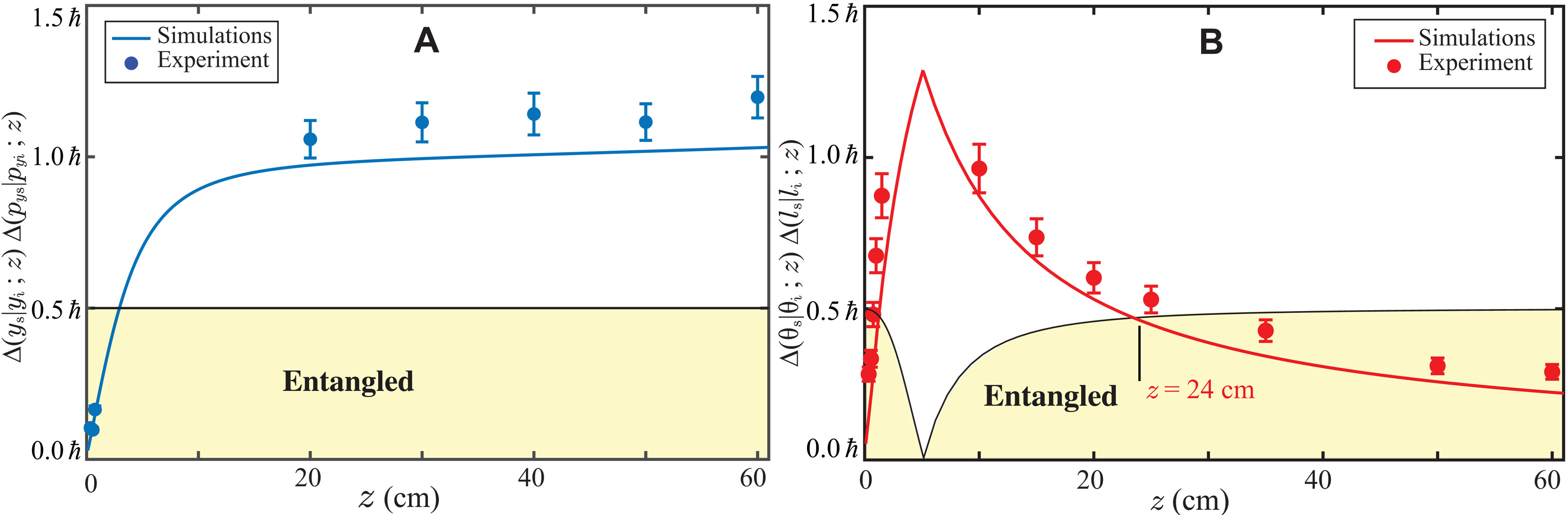}
    \caption{Propagation of entanglement in (A) position-momentum and (B) angle-OAM. The yellow region contains the entangled states. We can observe that while both types of entanglements are lost after a short propagation, the angle-OAM entanglement is recovered if the photons propagate further. Reproduced from Ref.~\cite{AbhiProagation} under the Creative Commons Attribution license (CC BY).}
    \label{F30}
\end{figure}

While the position JPD consistently broadens with propagation, the angle JPD broadens to a maximum and then reduces its width again. These behaviors have profound consequences in spatial entanglement, as shown in Figure~\ref{F30}. Position-momentum entanglement is shown in Figure~\ref{F30}(A), where the bottom yellow region contains entangled states. It is clear from the experiment that after a short propagation, position-momentum entanglement is lost. In Figure~\ref{F30}(B), the results for angle-OAM entanglement are presented. While the entanglement is lost, again after a short propagation, the entanglement is recovered again while the beam propagates further. These results can have important consequences in free-space entanglement distribution. 

\section{Applications in quantum imaging}\label{S8}
Spatial correlations, in particular, position/momentum correlations, have inspired numerous quantum imaging applications. In this section, we review how quantum correlations can be exploited for quantum imaging techniques and novel applications. 

Several reviews on advances in quantum imaging have recently been published~\cite{gilaberte_basset_perspectives_2019,defienne2024advances}. 
Since the first quantum imaging technique, ghost imaging, this field has flourished, introducing schemes that allow obtaining super-resolution~\cite{classen2017superresolution,unternahrer2018super,tenne2019super}, supersensitive~\cite{camphausen2021quantum,camphausen_fast_2023}, sub-shot noise imaging~\cite{brida2010experimental,samantaray2017realization}, holography \cite{defienne2021polarization,topfer2022quantum,topfer_synthetic_2025}, acquiring images with undetected light \cite{lemos2014quantum,fuenzalida2024nonlinear,gilaberte_basset_videorate_2021}, and distilling images from stray light \cite{defienne_quantum_2019,lloyd2008enhanced,gregory2020imaging,fuenzalida2023experimental}. Besides quantum imaging, spatially correlated photons have applications in entanglement distribution \cite{ortega2021experimental,achatz2023simultaneous}, quantum communications \cite{ecker_advances_2023,wei_towards_2022,ortega_implementation_2024}, quantum state engineering, quantum walks \cite{Hugospdcshaping,valencia2020unscrambling, grafe_integrated_2016, di_giuseppe_einstein-podolsky-rosen_2013,grafe_correlations_2013}, and adaptive imaging \cite{cameron2024adaptive}. 

\subsection{Ghost imaging}\label{S8A1}
\begin{figure}
                \centering
                \includegraphics[width=0.45\textwidth]{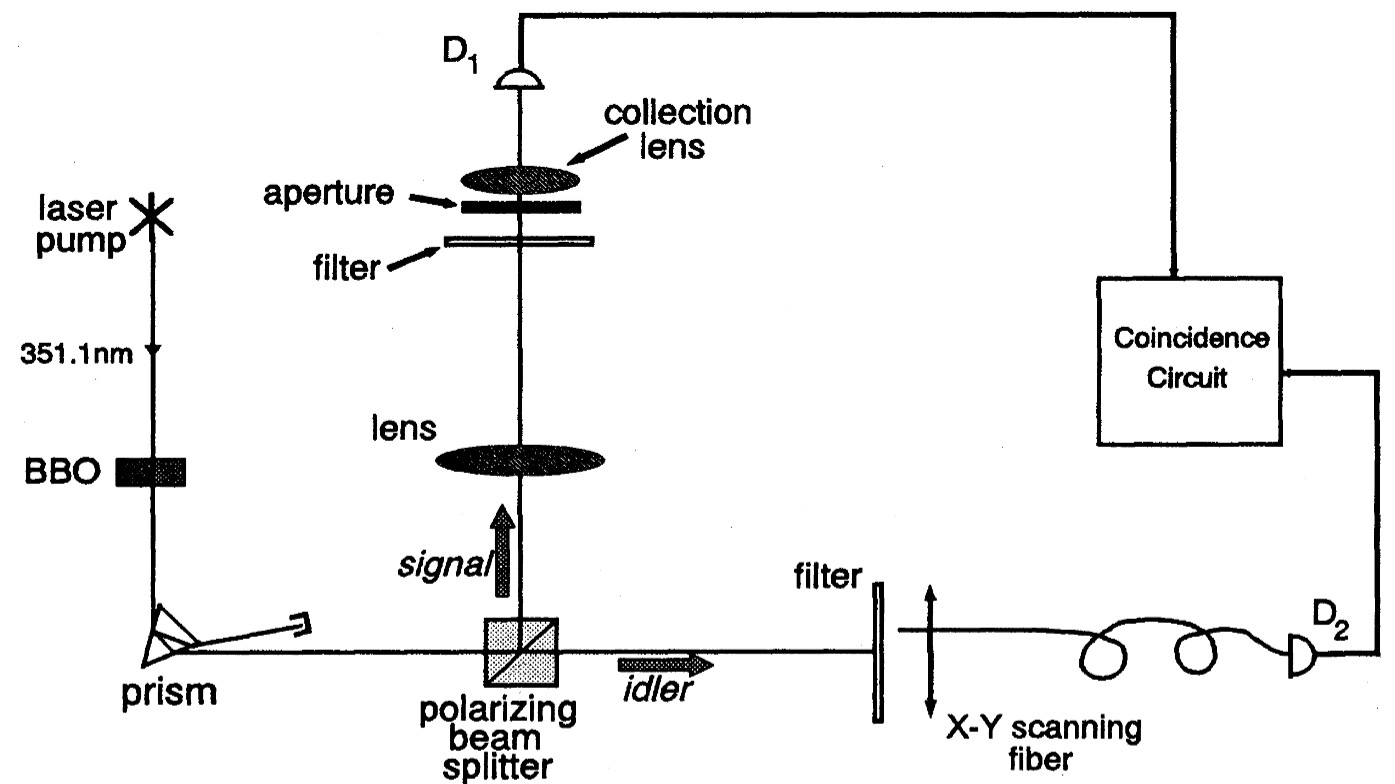}
                \caption{First ghost imaging setup. In current implementations, the X-Y scanning is replaced with a camera. Reproduced with permission. \cite{pittman_optical_1995} Copyright © 1995, American Physical Society.}
                \label{F32}
            \end{figure}

\begin{figure}
                \centering
                \includegraphics[width=0.45\textwidth]{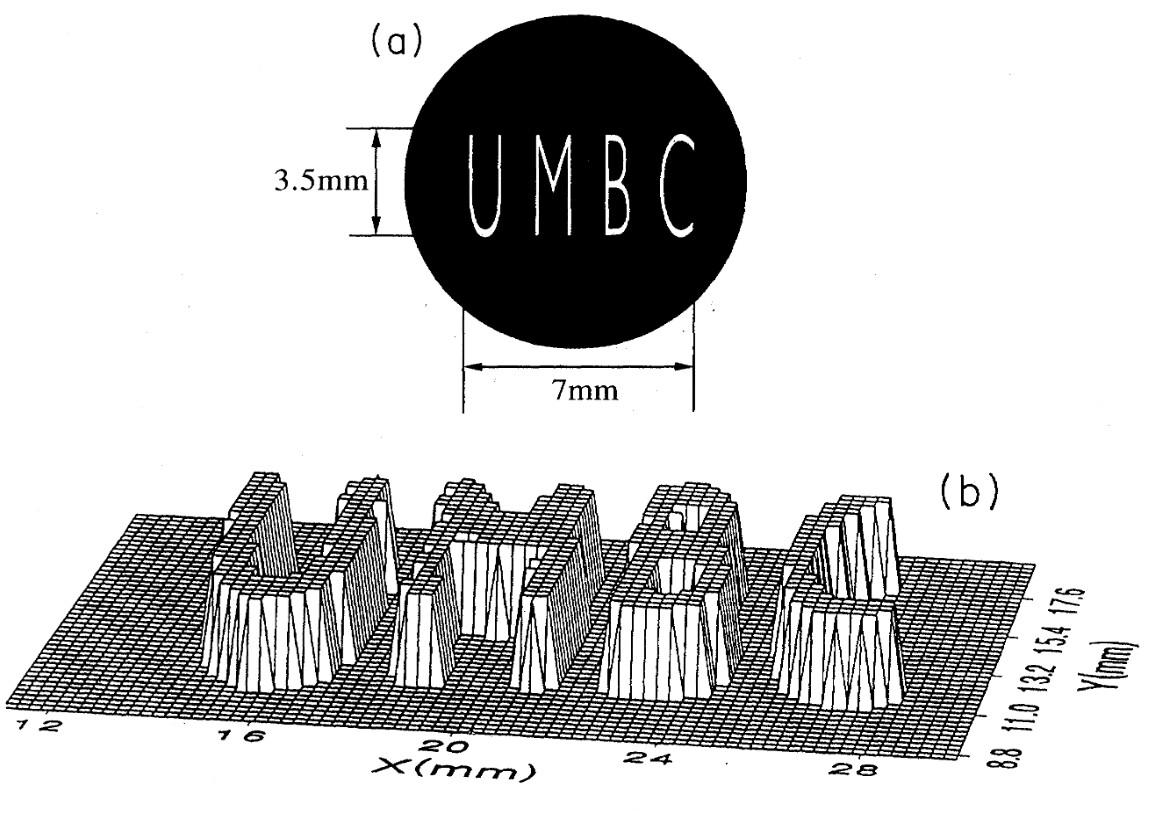}
                \caption{(a)  Object. (b) Coincidence counts cut to the half maximum value depending on the fiber tip's position in the transverse plane with a scanning step size of \qty{0.25}{\mm}. Reproduced with permission. \cite{pittman_optical_1995} Copyright © 1995, American Physical Society.}
                \label{F33}
            \end{figure}

\begin{figure}
\centering
\includegraphics[width=0.55\textwidth]{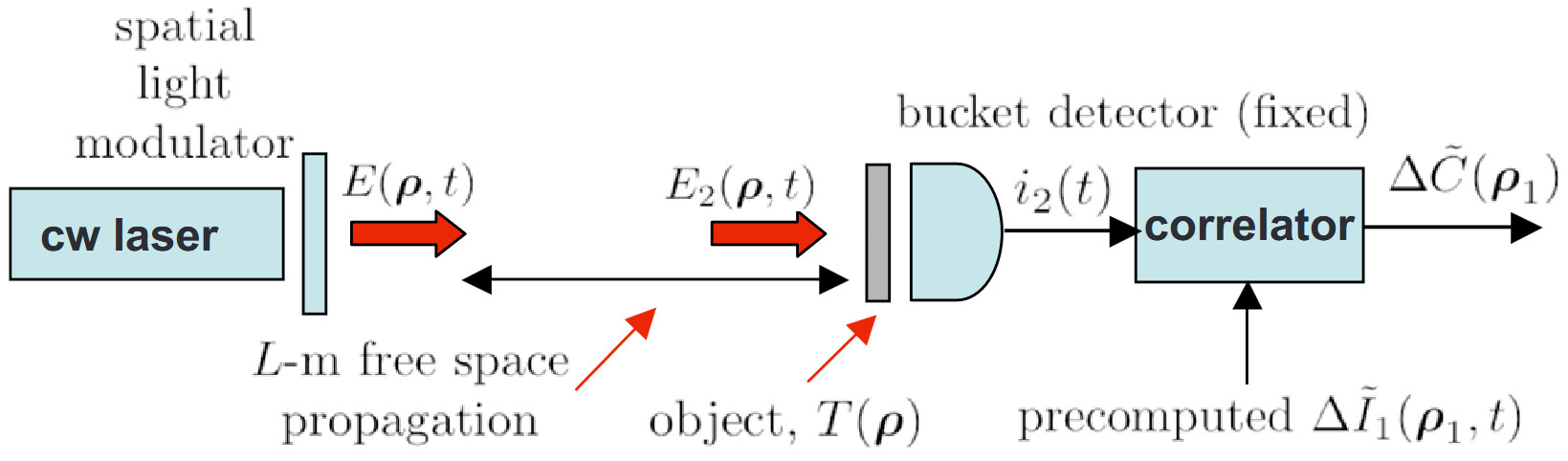}
\caption{Setup for computational GI. A spatial light modulator shapes the electric field $E$ before illuminating the target with transmission $T(\rho)$. The bucket detector measures the total intensity while a reference arm records the pattern, enabling image reconstruction. Reproduced with permission. \cite{shapiro_computational_2008} Copyright © 2008, American Physical Society.}
\label{F34}
\end{figure}

First introduced in 1995~\cite{strekalov_observation_1995, pittman_optical_1995}, (quantum) ghost imaging (GI) was the first imaging method to use spatial correlation of the photon pairs, generated by SPDC process. Based on coincidence measurement, one photon of the photon pair interacts with the object (aperture) before getting collected in a fiber while the other photon is collected in a scanning fiber (see Fig.~\ref{F32}) \cite{pittman_optical_1995}. Merging the information of coincident photon events and the position of the scanning fiber at that time allows for reconstructing the object in the signal path, resulting in the image shown in Fig.~\ref{F33}. Instead of a scanning fiber, one can also use a single-photon-sensitive camera \cite{aspden_epr-based_2013, fickler_real-time_2013, morris_imaging_2015}. Some other methods also enable probing the object with a different wavelength than the wavelength detected by the camera, allowing the use of more sensitive and less expensive cameras in the visible range while probing an object with a wavelength in the infrared spectrum or beyond~\cite{kim_infrared_2010, rubin_resolution_2008, karmakar_two-color_2010, aspden_photon-sparse_2015}. For example,  \cite{aspden_photon-sparse_2015} used a BBO-crystal to generate photon pairs at \qty{460}{\nm} and \qty{1550}{\nm}, illuminating the object with infrared light before triggering the measurement of the signal photon at an intensified camera with a CCD detector array (ICCD) \cite{aspden_photon-sparse_2015}. To compensate the electrical delay between heralding detector and camera, an image preserving free propagation path was introduced in the signal arm.\\
Contrary to the usual belief, GI does not necessarily rely on quantum correlations and can be achieved with classical correlation \cite{bennink_two-photon_2002, valencia_two-photon_2005}; this is known as classical GI. However, using quantum light possesses some advantages over its classical counterpart. \\ 
GI has obtained substantial improvement since its conception. On the one hand, it is also possible to get rid of the camera in GI using spatially structured light fields \cite{shapiro_computational_2008, bromberg_ghost_2009, moodley_all-digital_2023}. This can be implemented by inserting an SLM in the signal arm as shown in Figure~\ref{F34}. The SLM projects different patterns, each of which has a corresponding intensity profile. Thus, from this information, the object can be reconstructed~\cite{moreau_ghost_2018}. On the other hand, the acquisition time can be reduced with a reconstruction algorithm based on compressed sensing \cite{candes_introduction_2008, romberg_imaging_2008}. This algorithm utilizes redundancies in natural signals to decrease the number of measurements needed for the reconstruction of an object \cite{katz_compressive_2009}.

\subsection{Spatial resolution of correlation-based quantum imaging}\label{resolution-section}

The spatial resolution of GI is fundamentally limited by the photon pair's spatial correlations. In this review, we have noted that correlations can be measured in far-field or near-field planes, meaning that GI can employ either plane for image acquisition. Consequently, the image resolution is limited by different variables in these two planes. Another important imaging technique that uses spatial correlations is quantum imaging with undetected photons (QIUP)~\cite{lemos2014quantum}; however, in this technique, only one photon is detected while its partner remains undetected. Also, in QIUP, a nonlinear interferometer is employed for image reconstruction, with the image obtained from the interference terms. For more details about QIUP, the reader can check more dedicated reviews on this topic~\cite{fuenzalida2024nonlinear,barreto2022quantum,RevModPhys.94.025007}. In this section, we briefly point out the different regimes for the resolution of correlation-based quantum imaging techniques, such as GI and QIUP. 

Spatial resolution can be categorized into two different regimes: paraxial regime and non-paraxial regime. In the former, the resolution is mainly limited by the two-photon spatial correlations, while in the latter, the wavelength becomes increasingly relevant until it reaches an equivalent of the classical diffraction limit. We will explore the resolution of QIUP in these different regimes, since it has been expanded through all of them, but the reader can also take a look at works of resolution in GI~\cite{moreau2018resolution}.

Within the paraxial regime, the two-photon wavefunction can be described by two terms as shown in Equation~\ref{E8}, one part containing the pump's angular spectrum and the other containing the phase-matching condition. When thick crystals~\cite{gilaberte2023experimental} are used in quantum imaging, the resolution worsens due to the contribution of both terms. However, as we decrease the crystal length in comparison to $|\mathbf{q_i}|^{-1}$ and $|\mathbf{q_s}|^{-1}$, we reach the thin-crystal approximation~\cite{PhysRevA.57.3123}, which simplifies the treatment of resolution in the near and far field. The first work on the paraxial regime and within the thin crystal approximation was done by Fuenzalida \textit{et al.}~\cite{fuenzalida2022resolution} in the far-field plane, where the two-photon wavefunction can be reduced to the pump angular spectrum. The main findings are: (1) a bigger pump waist increases the resolution, and (2) the illumination wavelength settles the resolution. From here on, we consider the idler as the probing photon and the signal as the detected photon in a QIUP scheme. Hence, the momentum correlation width determining the resolution is given by

\begin{equation}
    \sigma_{FF}=\frac{f_i \, \lambda_i }{ \sqrt{2} \pi \, w_p},
\end{equation} 
where $f_i$ is the lens's focal length that produces the idler photon's far field, $\lambda_i$ is the idler wavelength, and $w_p$ is the pump waist. \\
The second work on the paraxial regime and the thin-crystal approximation is in the near-field plane, first theoretically by~\cite{viswanathan2021resolution} and later experimentally by~\cite{gilaberte2023experimental}. In this case, the two-photon wavefunction is reduced to the phase-matching condition, and consequently, the resolution depends on both SPDC wavelengths and the crystal length. Considering a magnification from the crystal to the object equal to 1, then the position correlation width is given by
\begin{equation}
    \sigma_{NF}=\frac{ \sqrt{L (\lambda_i+\lambda_s )}}{ 2 \sqrt{\pi}}.
\end{equation} 
The work done by Gilaberte \textit{et al.}~\cite{gilaberte2023experimental} also explores the limits of the paraxial regime, extending the work towards thick crystals by reducing the pump beam waist. The authors define the limit to be within the thin crystal as $w^2_p \gg [ \lambda^2_s L /( \lambda_s+\lambda_i )]$.

\begin{figure}[hbtp]
                \centering
                \includegraphics[width=0.45\textwidth]{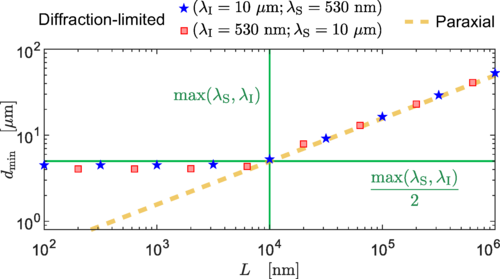}
                \caption{The resolution is limited by different variables within the paraxial regime. By reducing the crystal length at the scale of the SPDC wavelengths, this limit is given by $\max(\lambda_i,\lambda_s)/2$. Beyond the paraxial limit, the resolution is limited by the diffraction limit as $\max(\lambda_i,\lambda_s)$. Reproduced from Ref.~\cite{vega2022fundamental} under the Creative Commons Attribution (CC BY) license.}
                \label{vega}
            \end{figure}

Finally, the work of Vega \textit{et al.}~\cite{vega2022fundamental} explores the resolution limit with these types of techniques, reaching the non-paraxial regime. It is worth noting that this regime is accessible only when the crystal length is equal to/smaller than the SPDC wavelengths. Some of the main results are depicted in Figure~\ref{vega}. The figure shows that within the paraxial regime, the resolution is limited by $\max(\lambda_i,\lambda_s)/2$. If the crystal length continues decreasing, the theory no longer follows the paraxial regime but reaches the diffraction limit, where the new resolution limits are given by $\max(\lambda_i,\lambda_s)$.

For completeness, Santos \textit{et al.} have shown theoretically that  it is possible to achieve super-resolution in QIUP; more details in~\cite{elkin2022superresolution}.\\
    
\subsection{Quantum image distillation}\label{S8B}

Classical imaging is vulnerable to stray light, which reduces the quality of the image to the point that it cannot be visualized correctly. Quantum imaging offers a solution to this problem.

\begin{figure}[hbtp]
\centering
\includegraphics[width=0.45\textwidth]{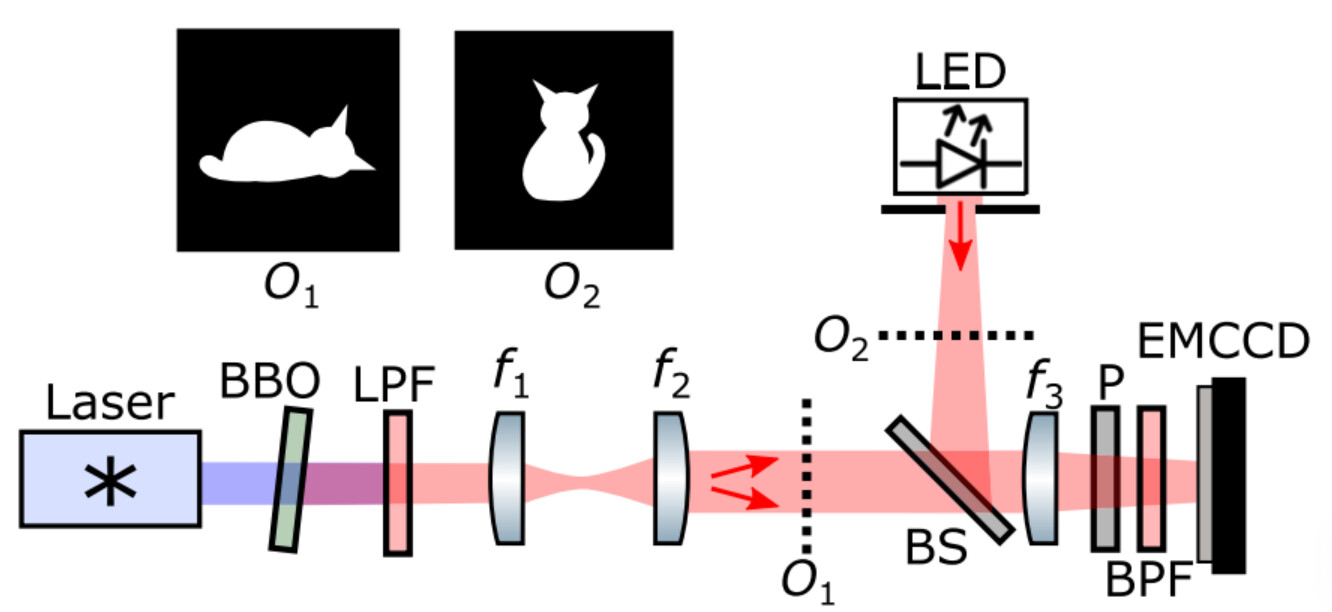}
\caption{Setup for quantum image distillation. Quantum image distillation setup: objects $O_1$ and $O_2$ are illuminated with quantum and classical light, respectively, and the two images are detected on an EMCCD camera. Reproduced from Ref.~\cite{defienne_quantum_2019} under the Creative Commons Attribution (CC BY) license.
}
\label{F35}
\end{figure}

\begin{figure}[hbtp]
\centering
\includegraphics[width=0.45\textwidth]{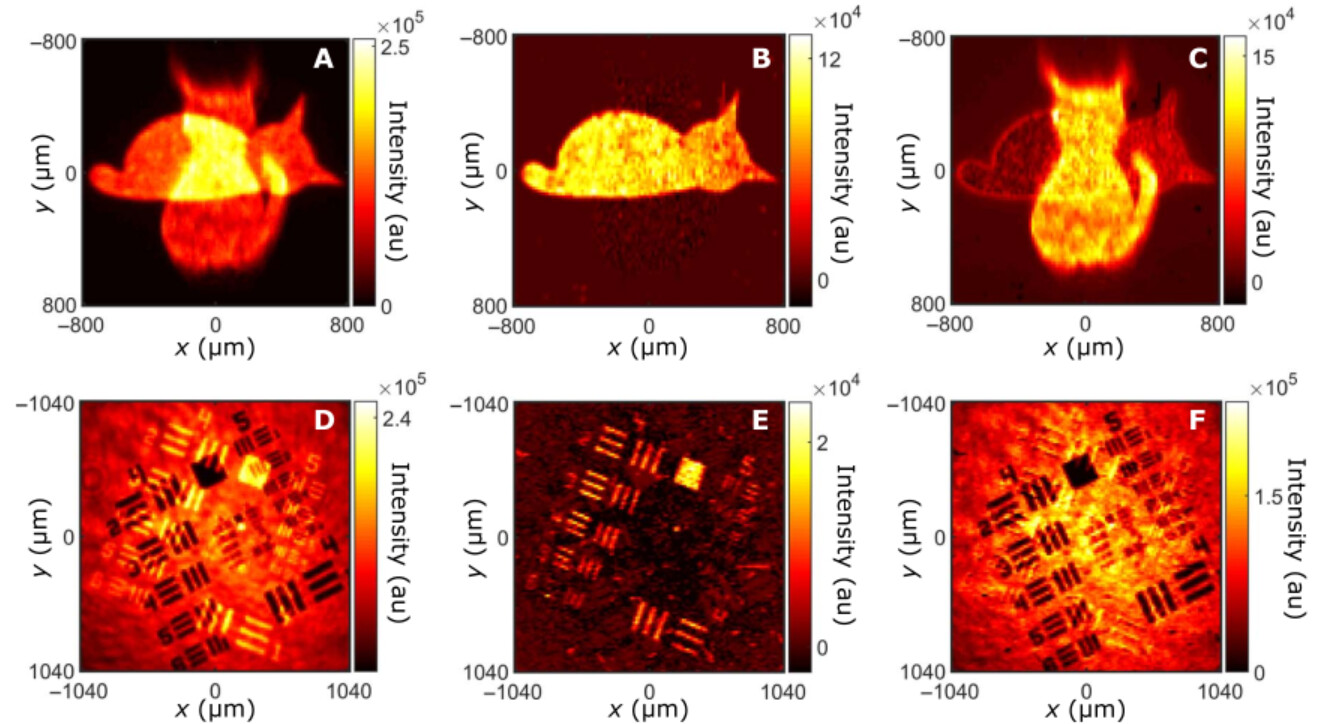}
\caption{(A) Direct intensity image taken by the camera. (B) Image produced by measuring the correlation function, revealing the object in the quantum arm. (C) Image of the object in the classical arm obtained by subtracting the correlation image from the direct intensity image. Reproduced from Ref.~\cite{defienne_quantum_2019} under the Creative Commons Attribution (CC BY) license.}
\label{F36}
\end{figure}

Quantum image distillation~\cite{defienne_quantum_2019} is a technique that uses quantum correlations to separate a quantum image from a source of noise, such as stray light. The technique leverages the distinct statistics of the quantum signal (sub-Poissonian) and the noise (Poissonian or super-Poissonian). The setup for this technique is shown in Figure~\ref{F35}. In one arm, spatially correlated photon pairs are generated in a $\beta$-barium borate (BBO) crystal which illuminates the object $O_1$ (sleeping cat) using two lenses $f_1$ and $f_2$. In another arm, classical light passes through an object $O_2$ (standing cat). Both images are superimposed by an unbalanced beam splitter (92:8) and imaged on an EMCCD camera using a lens $f_3$, followed by a polarizer and a bandpass filter. As one can see in Figure~\ref{F36}(a), both objects are overlapped in the intensity image. Retrieving the photon pair correlations, the quantum image alone can be made visible in Figure~\ref{F36}(b), or by subtracting it from the direct intensity image, one can get the pure classical image in Figure~\ref{F36}(c).
However, in the latter image, there is residual quantum light. This corresponds to the case when one photon of the SPDC pair is absorbed by the edges of the object or scattered to a different location.

\begin{figure}[hbtp]
\centering
\includegraphics[width=0.45\textwidth]{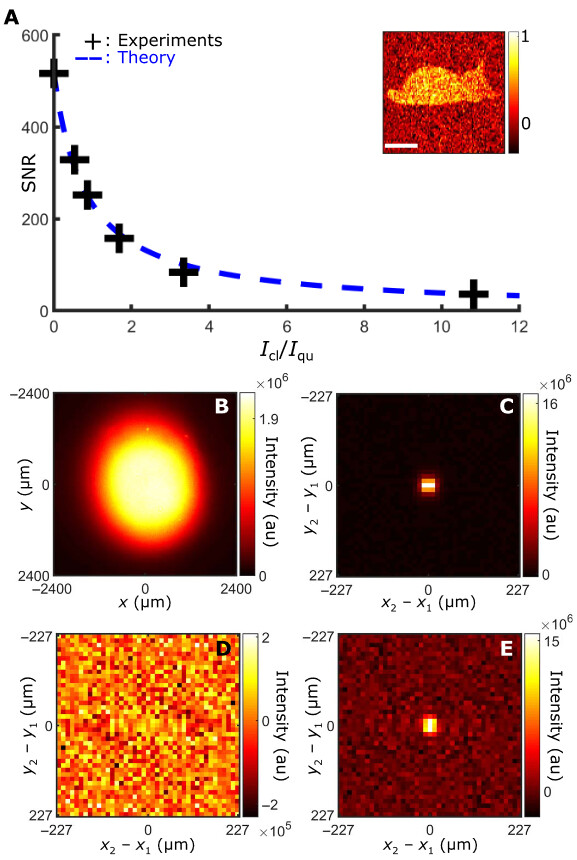}
\caption{(A) SNR measurements against classical-to-quantum intensity ratios. The crosses are experimental points, while the dotted line is the theoretical expectation. (B) intensity image with both classical and quantum light. (C)–(E) negative sum-coordinate projections of the JPD at $I_{cl}/I_{qu}=0$, $\infty$, and $11$, respectively. Reproduced from Ref.~\cite{defienne_quantum_2019} under the Creative Commons Attribution (CC BY) license.}
\label{F36.2}
\end{figure}

A study on the signal-to-noise (SNR) ratios is presented in Figure~\ref {F36.2}. Figure~\ref{F36.2}(A) presents the experimental values of the SNR against the ratios between classical and quantum illumination ($I_R=I_{cl}/I_{qu}$). It shows that the SNR decreases as $I_R$ increases. Figure~\ref{F36.2}(B) shows the sum of the intensities on the EMCCD sensor. The SNR is obtained from the minus-coordinate projection of $\Gamma(r,r)$, which represents the probability of coincidence of two pairs of photons at pixels separated by distance $r_1$-$r_2$. Figures~\ref{F36.2}(B)-(E) show the minus-coordinate projections acquired for $I_R$ of $0$, $+ \infty$, and 11, respectively.
    
The concept of distillation has a long-standing presence in the scientific community~\cite{bennet1996purification,pan2001entanglement}; however, these initial works were more focused on quantum communications. Furthermore, the term \textit{quantum illumination} introduced by Lloyd~\cite{lloyd2008enhanced}, and Tan \textit{et al.}~\cite{tan2008quantum}, also shares some similarities, but it is conceptually closer to sensing and metrology than to imaging. 

Finally, after the work of Defienne \textit{et al.}~\cite{defienne_quantum_2019}, several other approaches for quantum image distillation have been introduced ~\cite{zhang_experimental_2015,defienne_full-field_2021,gregory2020imaging,fuenzalida2023experimental,england2019quantum,zhang2020multidimensional}, showing the versatility of this technique.

\par

\subsection{Pixel super-resolution imaging}\label{S8A2}
        
Pixel super-resolution imaging~\cite{Defienne2022PixelSuperResolution} refers to the method that surpasses the normal pixel size limit of a detector by leveraging the fine-scale information available in spatially correlated photon pairs. In their work, Defienne \textit{et al.} demonstrated a two-fold increase in pixel resolution. 

The experimental setup is shown in Figure~\ref{F38}(a), where entangled photons are generated using a type-I BBO crystal. The crystal plane is imaged onto an object, which is a square-modulation amplitude grating. Then the object is imaged on the camera with a pixel pitch of 32~$\mu$m, and where the estimated correlation width at the camera plane is $\sigma_-$=13~$\mu$m.

\par
While most of the photon pair imaging schemes use the diagonal component of JPD to reconstruct an image's object, i.e., $\Gamma_{i j i j}=\left|t\left(x_i, y_j\right)\right|^4$ \cite{defienne_quantum_2019,Reichert2017,reichert_massively_2018,toninelli_resolution-enhanced_2019}, there is not a real advantage in resolution. This is due to the Moire pattern being present in the retrieved images. This is proven in Figure~\ref{F38} where an image is retrieved with two types of illumination: (b) direct intensity, and (c) two-photon JPD. These two images show identical resolutions.

\par
Nevertheless, it is possible to employ a different mechanism to improve the resolution of two-photon illumination. To do this, one can project the JPD along its sum-coordinate axis 
\begin{equation}
P_{i^{+} j^{+}}^{+}=\sum_{i=1}^{N_X} \sum_{j=1}^{N_Y} \Gamma_{\left(i^{+}-i\right)\left(j^{+}-j\right) i j},
\label{E42}
\end{equation}

where $N_X \times N_Y$ is the number of pixels of the illuminated region, $P^+$ is the sum-coordinate projection of the JPD and ($i^{+},j^{+}$) are sum-coordinate pixel indices. The resulting image of using this new projection is shown in Figure~\ref{F38}(f) for the same object. One can see that the spurious low-frequency modulation has been removed, and the resolution has increased by a factor of two. This high-resolution image is very similar to a direct intensity image acquired using a camera with a \qty{16}{\micro \m} pixel pitch.

Furthermore, spatial correlations can be harnessed to surpass Rayleigh’s diffraction limit. In Ref.~\cite{Grenapin2023Superresolution}, the authors demonstrated that \textit{biphoton Spatial-Mode Demultiplexing} (biphoton SPADE) significantly enhances super-resolution capabilities by leveraging the spatial correlations, or entanglement, inherent in two-photon light fields generated via SPDC. This enhancement is precisely quantified by the Schmidt number ($K$), which measures the strength of spatial entanglement between the photon pairs. The core advantage is derived from the increase in Fisher information (FI) associated with the coincidence measurements, as the FI is enhanced by a factor of $\sqrt{K}$ compared to ordinary SPADE techniques. Ordinary SPADE corresponds to the separable case where K=1, yielding a lower limit on FI. Since the total Fisher information scales as $FI_{tot} =\sqrt{K}$/2, any spatially entangled source ($K>$1) provides a measurable advantage, meaning the precision in estimating near-zero separations is limited only by the strength of the spatial correlations.\\

\begin{figure}
\centering
\includegraphics[width=0.45\textwidth]{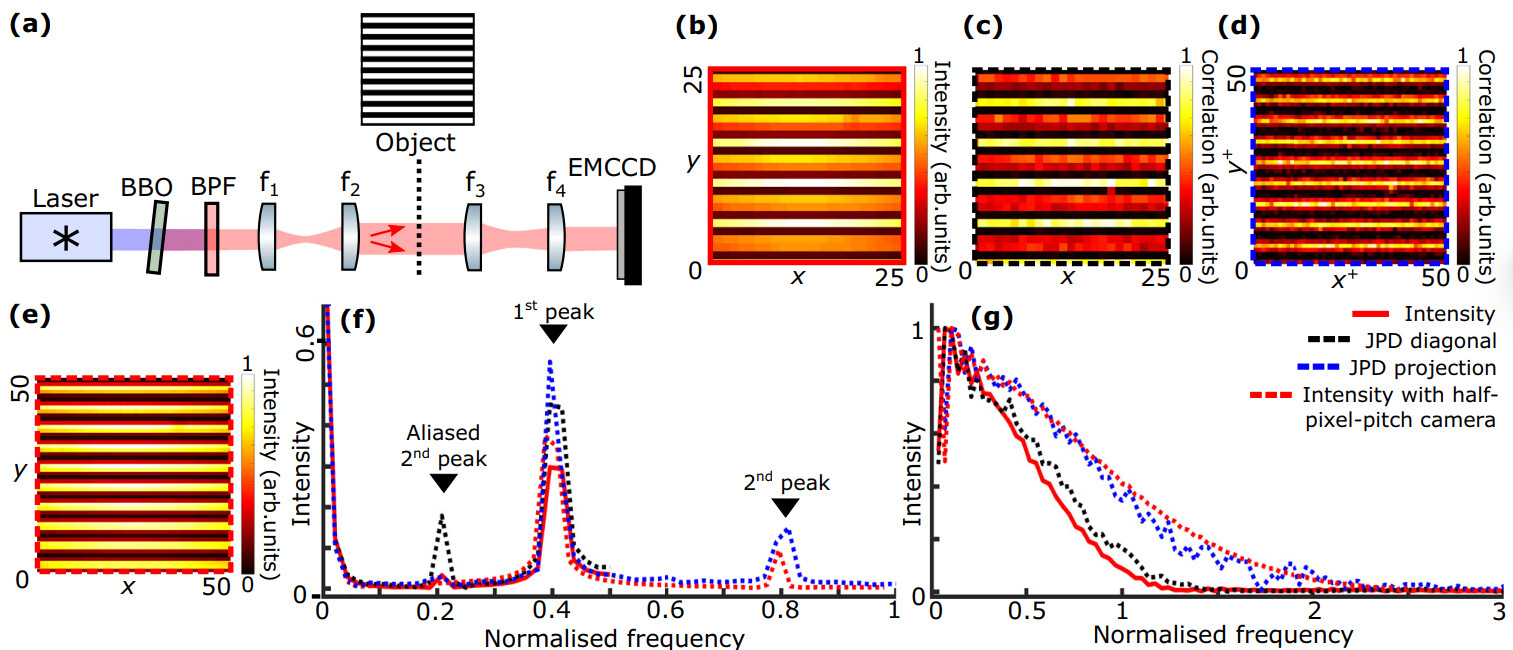}
\caption{(a) Experimental setup for pixel super-resolution. (b) Intensity image, (c) JPD diagonal image and (d) sum-coordinate projection of the JPD with a pixel pitch of $\sigma_-= \qty{32}{\mu \m}$. (e) Intensity image where a camera with half the pixel pitch is used compared to (b). (f) Line profile and (g) system modulation transfer function with the intensity image (solid red), diagonal image (dashed black), JPD sum-coordinate image (dashed blue), and intensity image acquired with a \qty{16}{\mu \m} pixel-pitch camera (dashed red). All frequencies are normalized to the reference frequency $1/\Delta$. Reproduced from Ref.~\cite{Defienne2022PixelSuperResolution} under the Creative Commons Attribution (CC BY) license.}
\label{F38}
\end{figure}
            
\subsection{Hiding images in quantum correlations}\label{S8A4}
Hiding images in quantum correlations is a technique where an object’s information is only concealed in the joint detection of photon pairs, rendering it invisible to direct intensity detection. 

The transfer of the pump's angular spectrum to the SPDC photon pairs~\cite{PhysRevA.57.3123}, allows to encode the information of an object illuminated by a pump to momentum correlation, i.e., the image can be retrieved just by measuring two-photon correlations in the far-field plane. Recently, Vernière and Defienne~\cite{verniere_hiding_2024} presented a more generalized study of the transfer of coherence and phase of the pump beam to the SPDC photon pairs. The experimental setup is shown in Figure~\ref{F38}(a). In the setup, a transmissive object (standing cat) was inserted in the pump beam and Fourier imaged on the crystal. 
Then, the crystal plane is Fourier imaged onto an EMCCD camera by the lens $f^\prime$. The direct intensity image of the crystal at far-field is a uniform SPDC output as shown in Figure~\ref{F38}(b). On the contrary, by examining the two-photon correlations, the standing cat is revealed, as shown in Figure~\ref{F38}(c). 

The authors also presented how a complex (transmission and phase) object can be hidden in the correlations, a relevant discussion about the lens distances, and a study of the trade-off between the SNR and the resolution---all these details in Ref.~\cite{verniere_hiding_2024}. 

\begin{figure}[hbtp]
\centering
\includegraphics[width=0.45\textwidth,]{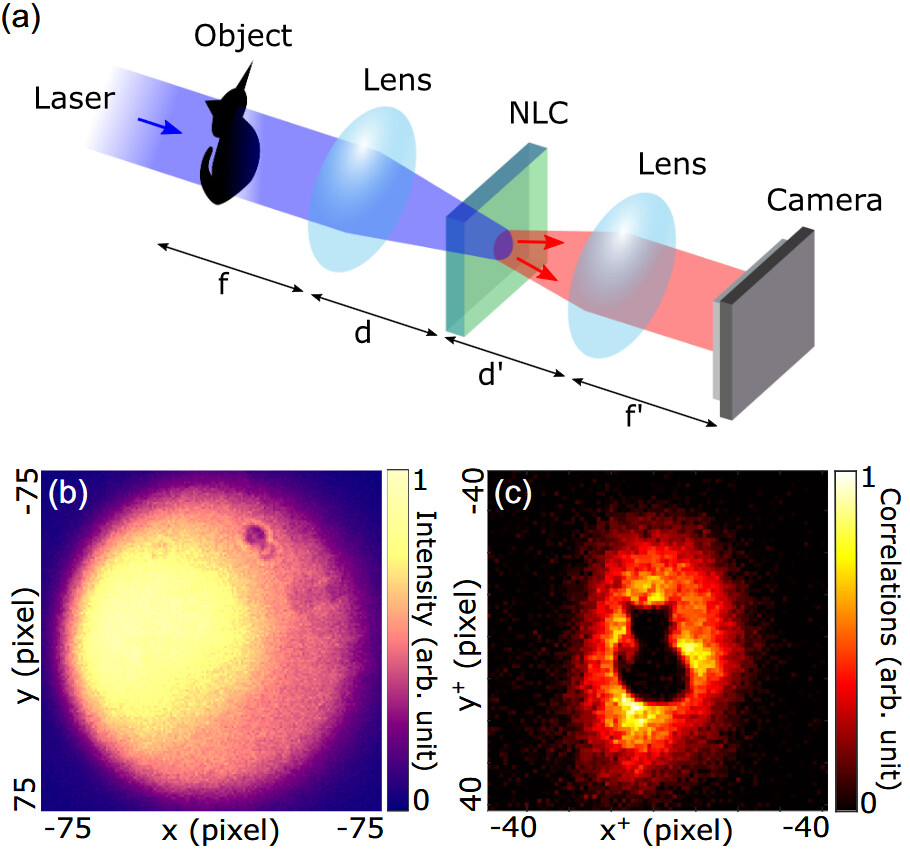}
\caption{(a) Experimental scheme for hiding an image in quantum correlations. A transmissive object (standing cat) is placed in the pump beam and Fourier imaged onto a thin nonlinear crystal (NLC), where SPDC photon pairs are generated. The crystal plane is then Fourier imaged onto an EMCCD camera. (b) Direct intensity measurement of the SPDC field in the far field, showing a uniform intensity distribution with no visible imprint of the object. (c) Two-photon correlation measurement in the far field, revealing the hidden image of the object encoded in the momentum correlations of the photon pairs. Reproduced with permission. \cite{verniere_hiding_2024} Copyright © 2024, American Physical Society.
}
\label{F38}
\end{figure}

    
\subsection{Quantum imaging in scattering environments}

Quantum imaging techniques based on SPDC have demonstrated remarkable resilience and precision in challenging optical conditions, where conventional imaging fails. This includes media that strongly scatter or blur light, such as frosted glass, biological tissue, or fog. 

\par
Early demonstrations~\cite{Gnatiessoro2019Imaging}, revealed that SPDC spatial correlations persist after the transmission through a thin scattering layer. This result established that quantum correlations can survive moderate diffusion, allowing for the reconstruction of spatial information beyond what is possible with classical light. Building on this foundation, Courme \textit{et al.}~\cite{Courme2023HighDimensional} employed wavefront-shaping and transmission-matrix techniques to transmit high-dimensional spatially entangled photon pairs through complex media, certifying entanglement 
after the scattering. Furthermore, Shekel \textit{et al.}~\cite{Shekel2024Shaping} demonstrated an advanced-wave feedback approach to correct the scattering on SPDC photons. 
A complementary work by Zhang \textit{et al.}~\cite{Zhang2024QuantumImagingBIO} introduced an imaging technique of biological samples using spatially and polarization-entangled photons, showing that quantum illumination can retrieve images from delicate, strongly scattering biological structures while minimizing sample exposure.

These studies highlight that spatially entangled photon pairs from SPDC sources provide a powerful tool for quantum imaging in scattering environments. Their resilience to optical noise, ability to maintain non-classical correlations, and compatibility with wavefront control techniques are of paramount importance for advanced imaging and sensing technologies, including biomedical diagnostics, remote sensing, and optical metrology under adverse conditions. Moreover, the robustness of spatial entanglement can find interesting applications in other quantum technologies, as we show next.\\

\par 

\section{Additional quantum technological applications}
Beyond imaging, spatial correlations have been applied in other quantum technological fields. This section covers some of those applications in areas such as quantum communications, quantum information, and quantum metrology.

\subsection{Quantum communications and information processing}

Quantum communication and information processing are mature fields that have impulsed the birth of several companies worldwide. While quantum communication uses quantum states for secure communication between parties, quantum information processing focuses on the manipulation, generation, distribution, and measurement of those quantum states. In this way, both fields are strongly related.

\begin{figure}
\centering
\includegraphics[width=0.45 \textwidth]{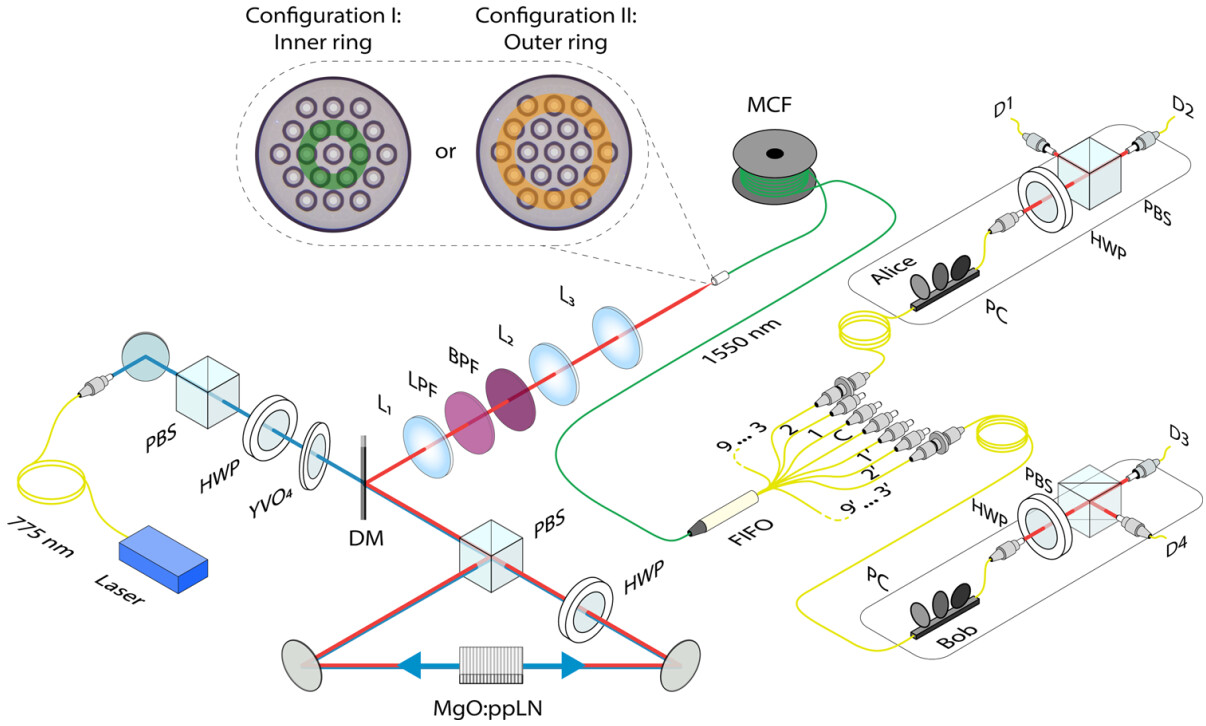}
\caption{Experimental setup for space-division multiplexing in quantum communications using spatial correlations in a multicore fiber (MCF). Polarization-entangled photon pairs are generated in a MgO:ppLN crystal in a Sagnac configuration and coupled into antipodal cores of the MCF by exploiting transverse momentum anti-correlations. A fan-in/fan-out device separates the individual cores into independent channels, which are subsequently analyzed for polarization correlations. Inset: cross section of a 19-core MCF showing the inner and outer ring configurations. Reproduced from Ref.~\cite{ortega_implementation_2024} under the Creative Commons Attribution (CC BY) license.}
\label{F42}
\end{figure}

Through this review, we have emphasized that momentum correlations are quantified with a two-photon joint detection. However, there are techniques based on quantum interference that only require the measurement of one of the photons for the same purpose. In the article by Hochrainer~\textit{et al.}~\cite{doi:10.1073/pnas.1620979114}, the two-photon momentum correlation is quantified through the effect of induced coherence without induced emission~\cite{PhysRevLett.67.318}. In this effect, the two-photon information is transferred to one of them as coherence, allowing for the measurement, among other properties, of the transverse momentum. Among other works, spatial correlations have found interesting applications in quantum entanglement distribution through multicore fibers (MCF). An MCF is a fiber that contains several cores distributed symmetrically within the same cladding, as shown in Figure~\ref{F42} inset for a 19-core MCF, where the cores are arranged into two rings. Therefore, this symmetry encourages the use of spatial correlations to distribute quantum states between antipodal cores. In a first experiment~\cite{ortega2022spatial}, photon pairs are generated in a MgO:ppLN crystal in a Sagnac configuration, resulting in polarization-entangled photon pairs. After filtering the pump, signal and idler photons are coupled into opposite cores of the MCF, exploiting the momentum anti-correlation. A fan-in/fan-out (FIFO) separates each fiber of the MCF into an independent channel, which is then analyzed in terms of polarization. Thus, this technique, known as \textit{space-division multiplexing}, enables the distribution of an entangled state through several parallel channels using a single entangled photon pair source. Later, Achatz \textit{et al.}~\cite{achatz2023simultaneous} achieved transmission of hyper-entanglement through an MCF, adding energy-time entanglement and path entanglement to the polarization one. This technique is also useful for quantum communications. By using space-division multiplexing, the two rings of the MCF (Figure~\ref{F42} inset, green circle: inner ring, orange circle: outer ring) can be used to perform quantum communications, where the key rate increases with the number of cores in the ring. In the experiment, a key rate of 7.3\,kbit/s and 34.5\,kbit/s was obtained in the inner and outer rings, respectively, with a quantum bit error rate of 3.8\% over a 24-hour measurement.

\begin{figure}
\centering
\includegraphics[width=0.45 \textwidth]{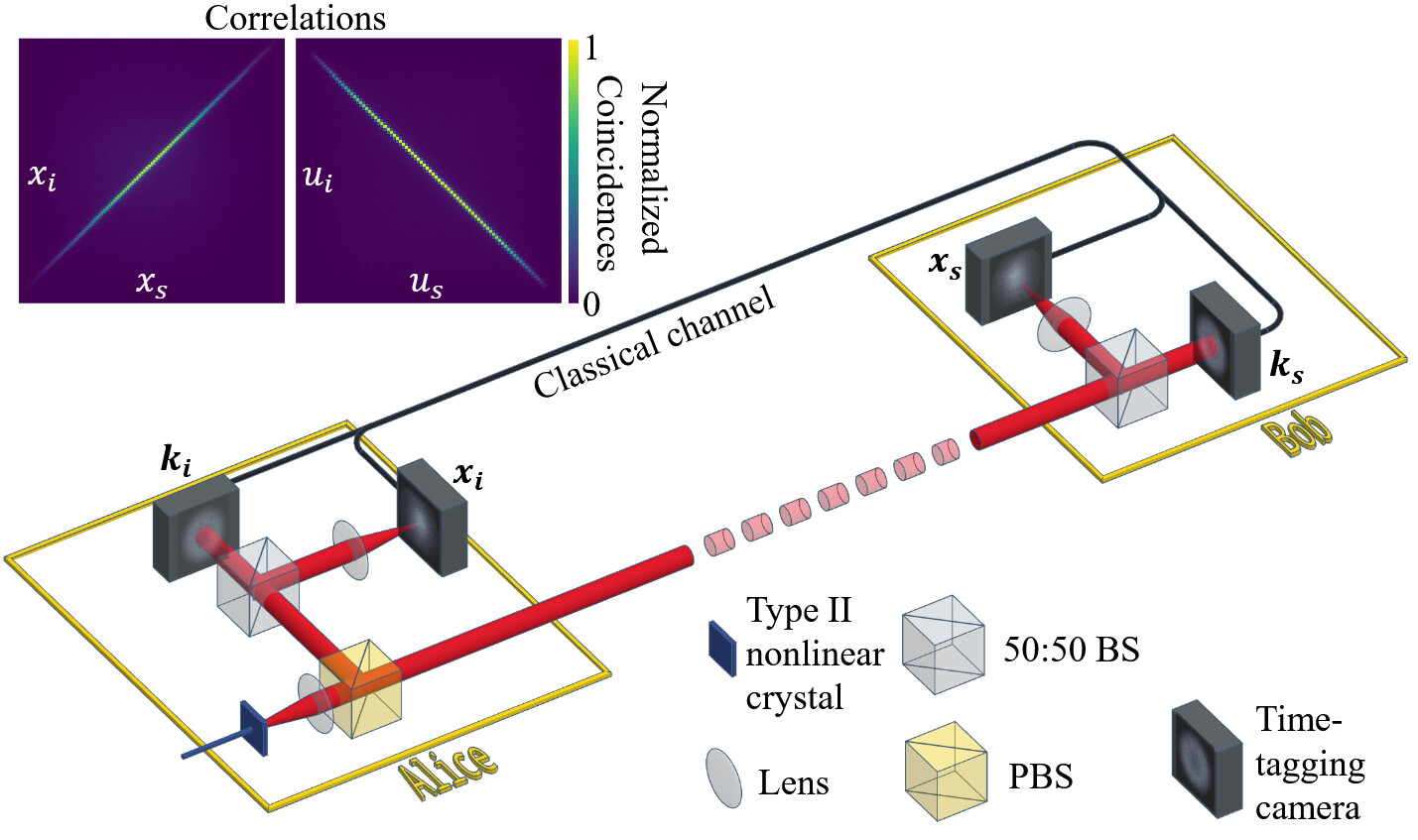}
\caption{Experimental setup for quantum key distribution based on position–momentum entanglement. Photon pairs generated via type-II SPDC are separated by a polarizing beam splitter and sent to two distant parties (Alice and Bob). Each party randomly measures the photon in either the position or momentum basis using a time-tagging single-photon camera. When the same basis is chosen, the measurement outcomes are ideally perfectly correlated, enabling high-dimensional quantum key distribution. 
Adapted from~\cite{scarfe_quantum_2025}.}
\label{F40}
\end{figure}

Quantum communications can also be implemented directly using the mutually unbiased bases for position and momentum~\cite{scarfe_quantum_2025}. The experimental setup is displayed in Figure~\ref{F40}. A photon pair generated via type-II SPDC is split by a polarizing beam splitter, where one photon is sent to Alice and the other to Bob. Each observer measures the photon in one of two mutually unbiased bases for position and momentum by a time-tagging single-photon camera. Whenever Alice and Bob use the same basis, their outcomes are ideally perfectly correlated. With this technique, the authors demonstrated a photon efficiency of \num{5.07} bits per photon with 90 spatial modes and a bit key rate of \qty{0.9}Kbit s$^{-1}$ using 361 modes. In addition, in Ref.~\cite{Bhat2025}, the authors propose and experimentally demonstrate a high-dimensional (HD) implementation of the BBM92 quantum key distribution protocol using position–momentum entanglement. A notable difference with the previous works is that 
the detection area is segmented into equiprobable macropixels using a \textit{fryum wheel} segmentation strategy designed to minimize crosstalk and reduce the quantum-dit error ratio. Importantly, the method discards events from ambiguous border pixels, a counterintuitive step that significantly improves the secure key rate. The authors obtained a secure key rate of 3.2$\pm$ 0.1 bits/photon. 
These schemes show the potential that can be achieved by exploiting the high dimensionality of spatial modes.

\subsection{Quantum metrology}
In the field of quantum metrology, some theoretical studies are exploring the benefits of using spatial correlations to surpass the shot-noise limit \cite{Lyons2016Precision} and the Heisenberg uncertainty limit \cite{DiLorenzo2013Correlations, Bullock2014Focusing}. Experimentally, Zhang \textit{et al.}~\cite{Zhang2025QuantumEnhanced} demonstrated a quantum-enhanced beam tracking that exceeds the conventional Heisenberg uncertainty limit by exploiting spatial entanglement of photon pairs. In their setup, similar to the one depicted in Figure~\ref{F40}, entangled photons produced via type-II SPDC exhibit strong correlations in both position and momentum. One photon served as a reference, while its partner was propagated through a channel that introduced controlled displacements and angular deflections. By performing joint coincidence measurements of position and momentum, the authors show how to surpass the Heisenberg uncertainty relation. 

\begin{figure}[hbtp]
                \centering
                \includegraphics[width=0.45\textwidth, trim=0pt 0pt 0pt 0pt, clip]{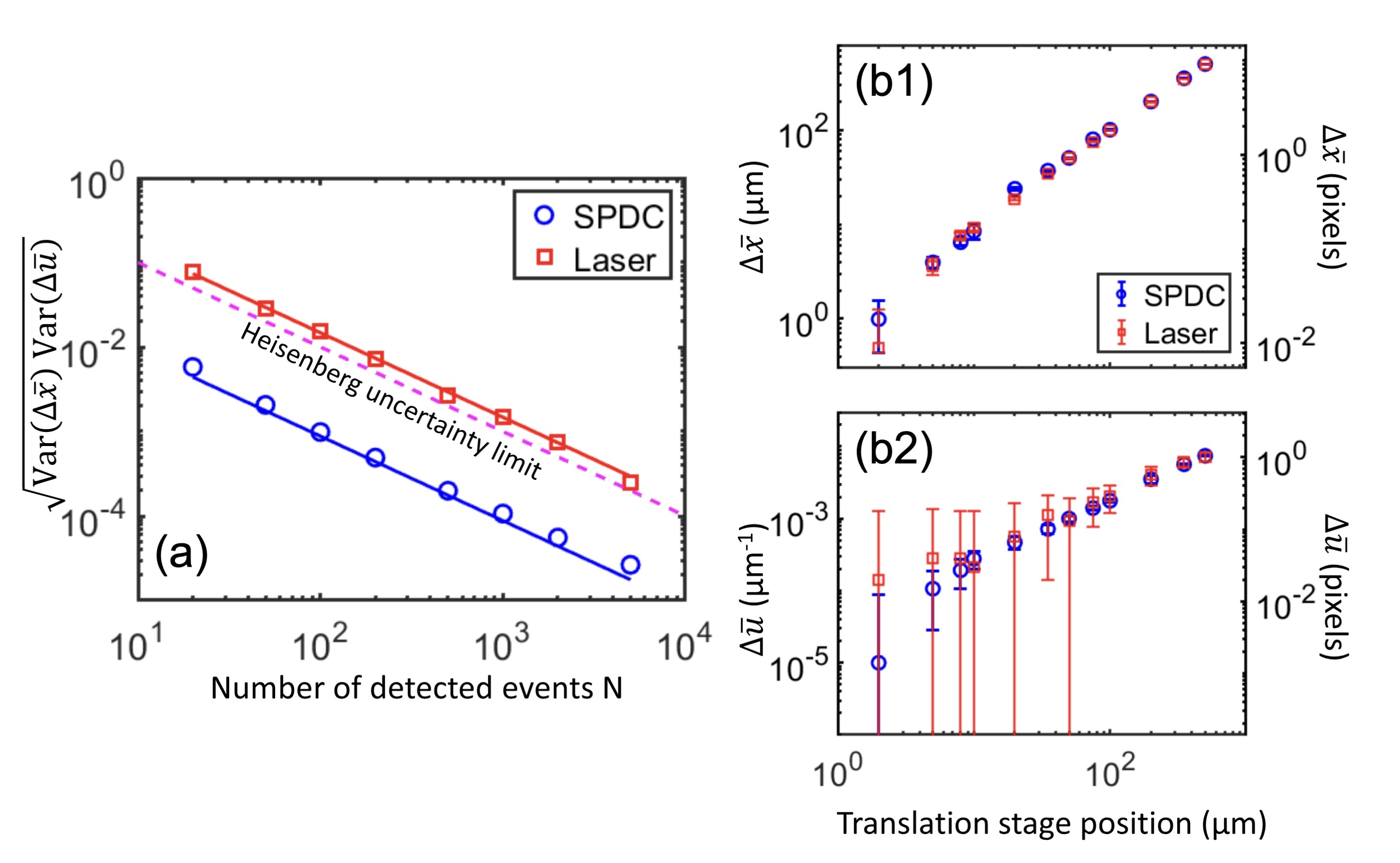}
                \caption{(a) Position-momentum uncertainty product for the change in beam trajectory as a function of the number of detected events. This is compared for the cases of using the spatial correlations of SPDC (blue circles) and a laser (red squares). The solid lines are the expected uncertainty product based on the measured correlation and beam widths. (b1), (b2) Compares the measured beam displacement in the position and momentum planes for $N=5000$ while using SPDC spatial correlations (blue circles) and a laser (red squares). Reproduced with permission. \cite{Zhang2025QuantumEnhanced} Copyright © 2025, American Physical Society.}
                \label{F101}
\end{figure}

Figure~\ref{F101} illustrates the experimental results. Figure~\ref{F101}(a) plots the uncertainty product for the beam-trajectory estimation as a function of the number of detected events for both photon pairs and single photons. The SPDC photons data lie below the dashed line marking the Heisenberg uncertainty limit (HUL), demonstrating a clear quantum advantage. Panel (b) consists of two sub-figures: (b1) shows measured displacements in the near field (position plane) for various mirror translations, while (b2) shows measurements in the far field (momentum plane). Although both sources perform similarly for position measurements, the SPDC-based scheme exhibits significantly smaller uncertainties in momentum estimation—highlighting the enhanced precision made possible by position–momentum entanglement.\\
This work represents a significant step forward in applying position–momentum entanglement to quantum metrology. Beyond its fundamental significance, it demonstrates that spatially entangled photons can enhance precision in practical sensing tasks such as beam tracking, even under strong background illumination. Although the current implementation relies on coincidence post-selection and controlled laboratory conditions, it paves the way toward real-time, unconditional quantum-enhanced tracking enabled by high-efficiency detectors. 


These applications underscore that spatial entanglement though sensitive to propagation and environmental factors, remains a powerful and versatile resource for advancing quantum communication, metrology, and emerging photonic technologies when carefully engineered and harnessed.\\

\section{Summary and Outlook}
Spatial entanglement in position and momentum has become one of the most versatile resources in quantum optics. Over the past few decades, the theoretical groundwork established by early studies on the Einstein–Podolsky–Rosen (EPR) paradox has led to numerous experimental advancements, confirming the genuine nonlocality of spatially entangled photon pairs generated by spontaneous parametric down-conversion (SPDC). From fundamental tests of quantum realism to application-oriented demonstrations such as quantum imaging, ghost imaging, and high-dimensional quantum communication protocols, the field has expanded in both scope and sophistication.

Beyond current demonstrations, the high-dimensionality structure of position–momentum entanglement positions it as a key resource for next-generation quantum communication protocols that demand large information capacity and intrinsic noise resilience. Various approaches have been introduced to quantify, certify, and fully characterize these correlations. Traditional slit-based coincidences provided some of the earliest evidence for EPR violations, while more recent techniques utilize cameras (EMCCD, SPAD arrays) that capture entire two-dimensional joint probability distributions in real-time. These camera-based techniques have significantly reduced acquisition times and opened the door to studies of dynamic or nonideal conditions, making spatial entanglement measurements increasingly robust and accessible.

Control over spatial coherence and transverse profile of the pump beam has emerged as a powerful method for tailoring the entanglement properties of SPDC photons. By tuning key parameters---crystal length, beam waist, wavelength, and phase-matching conditions---one can manage the position or momentum correlation widths, thereby shaping the trade-offs between resolution, brightness, and degree of entanglement. The ability to engineer tailored spatial correlations suggests promising applications in adaptive quantum imaging, and quantum-enhanced sensing. Research on partially coherent pumps and wavefront-shaping in SPDC underscores the potential for \textit{engineered entanglement}, allowing one to design custom correlations for specialized applications, from advanced holography to sensor fusion in turbulent environments.

Despite these achievements, challenges persist. Long-distance free-space transmission of spatially entangled photons, for instance, faces hurdles such as diffraction, turbulence, and scattering. Exploring the limits of entanglement transport and developing protocols for entanglement distillation or error correction will be essential steps in converting laboratory-scale demonstrations into robust quantum networks. Integration with waveguide photonics remains a promising frontier, as it can provide stability, scalability, and synergy with cutting-edge on-chip devices.

Looking ahead, the interplay of spatial entanglement, quantum metrology, and machine learning may lead to new protocols capable of extracting detailed sample information from minimal photon resources. Meanwhile, harnessing high-dimensional OAM-based or hybrid (polarization-plus-spatial) states will expand the capacity of future quantum communication channels. As integrated photonics matures, exploiting position–momentum entanglement across several waveguide modes (including controllable multimode waveguides)
could yield compact, scalable, and multifunctional quantum processors with capabilities not achievable using polarization-only qubits. With progresses, we anticipate more sophisticated entanglement sources, faster and more sensitive detection schemes, and increasingly advanced quantum-optical architectures designed around the unique strengths of position-momentum entanglement. Together, these advancements promise to deepen our fundamental understanding of nonlocality while accelerating the deployment of quantum technologies across a wide range of scientific and industrial fields.\\

\section{Conclusion}
This review article contains a comprehensive and unified overview of spatial entanglement in the position–momentum domain, tracing its development from foundational quantum mechanics to modern photonic technologies. By examining the generation of entangled photon pairs through SPDC, methods to measure and tailor their joint spatial distributions, and the influence of pump coherence and phase-matching conditions, we have covered how spatial correlations can be precisely engineered and exploited. We also surveyed a broad range of applications—from quantum imaging, ghost imaging, and imaging with undetected photons to pixel super-resolution, image distillation, and imaging in scattering environments—highlighting how spatial entanglement provides capabilities that surpass classical limits in resolution, noise robustness, and information extraction.
Furthermore, we emphasized emerging opportunities in quantum information science, including high-dimensional quantum communication, entanglement distribution through complex media and multicore fibers, and quantum-enhanced metrology. These advances underline that spatial entanglement is not only a fundamental manifestation of nonlocality, but also a versatile and high-dimensional resource for next-generation quantum technologies. As detection platforms improve and integrated photonics continues to mature, the ability to generate, control, and apply position–momentum entanglement will play an increasingly central role in scalable quantum systems, enabling more sophisticated protocols in sensing, communication, and computation.

\medskip
\section*{Acknowledgment}
The authors acknowledge financial support from the German Federal Ministry of Education and Research (BMBF) within the funding program “quantum technologies—from basic research to market” with contract number 13N16496 (QUANCER) and from the German Research Foundation (DFG) under the project number 552245798.\\

\section*{Conﬂict of Interest}
The authors declare no conﬂict of interest.\\
            
\section*{Keywords}
spatial entanglement, position-momentum entanglement, high-dimensional entanglement, quantum imaging, quantum communication, quantum information, quantum metrology \\

\end{document}